\documentclass[11pt]{article}
\usepackage{graphicx}
\usepackage{epsfig}
\usepackage{bbm}

\newcommand{\be}{\begin{eqnarray}}
\newcommand{\ee}{\end{eqnarray}}

\def\nue{{\nu_e}}
\def\anue{{\bar\nu_e}}
\def\numu{{\nu_{\mu}}}
\def\anumu{{\bar\nu_{\mu}}}

\newcommand{\ms}{\Delta m^2_{21}}
\newcommand{\ma}{\Delta m^2_{31}}
\newcommand{\meff}{\Delta m^2_{\rm eff}}

\newcommand{\sss}{\sin^2 \theta_{12}}
\newcommand{\sch}{\sin^2 \theta_{13}}
\newcommand{\stch}{\sin^2 2\theta_{13}}
\newcommand{\sa}{\sin^2 \theta_{23}}
\newcommand{\sta}{\sin^22 \theta_{23}}

\newcommand{\stcht}{\sin^2 2\theta_{13}{\mbox {(true)}}}
\newcommand{\sat}{\sin^2 \theta_{23}{\mbox {(true)}}}

\newcommand{\dcpt}{\delta_{CP}{\mbox {(true)}}}
\newcommand{\dcp}{\delta_{CP}}

\def\nn{\nonumber}
\def\ltap{\ \raisebox{-.4ex}{\rlap{$\sim$}} \raisebox{.4ex}{$<$}\ }
\def\gtap{\ \raisebox{-.4ex}{\rlap{$\sim$}} \raisebox{.4ex}{$>$}\ }

\newcommand{\oct}{{\rm octant~of~\theta_{23}}}

\def\gs{\mathrel{
   \rlap{\raise 0.511ex \hbox{$>$}}{\lower 0.511ex \hbox{$\sim$}}}}
\def\ls{\mathrel{
   \rlap{\raise 0.511ex \hbox{$<$}}{\lower 0.511ex \hbox{$\sim$}}}}
\newcommand{\bea}{\begin{equation} \begin{array}{c}}
\newcommand{\bead}{\begin{equation} \begin{array}{cccc}}
\newcommand{\eea}{ \end{array} \end{equation}}

\begin{document}

\title{\bf Determining the Octant of $\theta_{23}$ with PINGU, T2K, NO$\nu$A 
and Reactor Data
}
\author{
\\
Sandhya Choubey\thanks{email: \tt sandhya@hri.res.in},~
Anushree Ghosh\thanks{email:\tt anushree@hri.res.in}
\\
{\normalsize \it$^{a}$Harish-Chandra Research Institute, Chhatnag Road, Jhunsi, Allahabad 211 019, India}\\ \\
}

\maketitle
\begin{abstract}\noindent

We explore the prospects of determining the octant of $\theta_{23}$ 
with atmospheric neutrinos at PINGU. We study in detail the impact of 
energy and angle resolutions of the neutrino on the octant sensitivity. 
We show that the systematic uncertainties on the atmospheric 
neutrino flux predictions, especially the ones which affect the energy and zenith 
angle spectrum of the neutrinos, make a rather drastic reduction of the 
sensitivity of PINGU. We also study the 
prospects of measuring the octant of $\theta_{23}$ in the long baseline 
experiments T2K and NO$\nu$A in conjunction with the reactor experiments.
We study this for two configurations of NO$\nu$A and T2K and make a comparative 
analysis of them. From just 3 years of PINGU 
data, the octant could be determined at more than $3\sigma$ C.L. for 
$\sin^2\theta_{23}<0.419$ and $\sin^2\theta_{23}>0.586$ if we add the reactor data 
and if normal hierarchy is true. On addition of the data from T2K and NO$\nu$A, the 
sensitivity improves so that the octant could be determined at the $4\sigma$ C.L. for 
$\sin^2\theta_{23}<0.426$ and $\sin^2\theta_{23}>0.586$ if normal hierarchy is true. 
Even a $5\sigma$ significance for the right octant can be achieved if $\sat<0.413$ 
for the true normal hierarchy. The sensitivity for the true inverted hierarchy is lower and we 
expect a 
$3\sigma$ determination of $\oct$ for 
$\sat<0.43$ and $>0.585$ from the combined data set for this case.

\end{abstract}
\bigskip
\bigskip

\section{Introduction}

PINGU \cite{pingu} (Precision IceCube Next Generation Upgrade) 
has been proposed as a low energy extension of the 
already existing and successfully running IceCube 
detector \cite{icecube}. While the energy threshold of the full IceCube detector is 100 GeV, 
PINGU is envisioned to have an energy threshold of 
a few GeV, thereby allowing it to function as a
low energy 
atmospheric neutrino experiment, with an effective fiducial mass in the multi-megaton range. 
The plan is to increase the number of strings, increasing the optical module density, and hence  
increasing the photo-coverage of the region. This will reduce the energy threshold for the 
detection of the muons. The number of optical modules have already been 
increased for the existing IceCube Deep Core (ICDC) area of the detector which 
has an energy threshold of 10 GeV and which has already released its first data 
on atmospheric neutrinos with energies greater than 10 GeV \cite{icdc}. 
The ICDC has 8 extra strings in addition to the 7 
strings belonging to the original IceCube design, providing a string spacing of 
about 75m. 
The plan is to increase the string density by putting   
20 to 40 additional strings. 
The 20 string option would allow for 26m spacing between strings. 
The 40 string option will obviously have a larger 
effective area compared to the 20 string option and possibly a lower energy threshold, and 
hence will give better results.
The large fiducial mass gives PINGU an edge over other competitive experiments. 
The potential of measuring the neutrino mass hierarchy at PINGU has been studied by 
quite a few groups \cite{pingumh,smirnovmh,skamh,sr,ww,bsmh,hagiwara13}. 
\\

Amongst the other issues in neutrino physics that remain to be probed, is the 
determination of the octant of $\theta_{23}$ mixing angle, in case it is found to be 
different from maximal. Various ways have been suggested in the literature to 
determine the $\oct$ in the current and next generation neutrino oscillation 
experiments. One easy way is to combine the data from reactor experiments 
with the $\nue$-appearance data from conventional accelerator experiments 
\cite{minakataoctant,mahn,tenyrs09}.
The reactor experiments return a pure measurement of the mixing angle 
$\stch$, while the $\nue$-appearance data from conventional 
accelerator experiments measure the combination $\sa\stch$, at leading order. 
Using this combined analysis, one could then extract information on the $\oct$.
Another approach studied in the literature has been to combine the 
$\nue$-appearance channel in long baseline experiments with the $\numu$-disappearance 
channel \cite{t2k2.5,animesh}.
The upshot of this reasoning is that the best-fit $\theta_{23}$ preferred by the 
appearance channel is different from the best-fit $\theta_{23}$ favored 
by the disappearance channel. This generates a synergy between the 
two data sets in the long baseline experiment, leading to an octant 
sensitivity. 
The third way  is to use $\ms$ dependent 
terms in the oscillation probability, which depend on either $\sa$ 
or $\cos^2\theta_{23}$, leading to $\theta_{23}$ octant sensitivity. This was shown 
in the context of sub-GeV $\nue$-events from atmospheric neutrinos at 
a water Cerenkov experiment like Super-Kamiokande \cite{gms2004}
where the $\ms$ driven oscillatory terms bring in an $\oct$ dependence in the 
low energy electron event sample. 
Finally, one can use the $\oct$ dependence in the earth matter effects 
in the $P_{\mu\mu}$ channel to get a measure on this parameter \cite{octant}. 
Determining the $\oct$ through observation of earth matter effects in 
$P_{\mu\mu}$ channel has been studied for atmospheric neutrino 
experiments with  magnetized iron detectors  
\cite{octant,indumurthyoctant,ss10}, water Cerenkov detectors \cite{hk}
and liquid argon detectors \cite{animesh,gandhi2012,lartalk}. 
\\

In this paper, we mainly focus at the prospects of determining the $\oct$ through 
observation of earth matter effects in megaton-class Cerenkov detectors. In particular, 
we will work with the PINGU set-up, though our results are also valid for ORCA 
(Oscillation Research with Cosmics in the Abyss) \cite{orca}, 
with 
suitable adjustments for the detector specifications like effective volume, 
reconstruction efficiency and energy and angle resolutions. The physics potential 
of PINGU has been studied in \cite{pingumh,smirnovmh,skamh,sr,ww,bsmh,hagiwara13}.
In particular, the issue of determining the octant of $\theta_{23}$ was briefly discussed in 
\cite{smirnovmh}. Here, we will present detailed analysis of the reach of the PINGU 
experiment towards the determination of the $\oct$, taking into account the 
detector specifications provided by the PINGU collaboration, as well as after including 
full systematic uncertainties on the atmospheric neutrino fluxes. We will show in detail how the 
projected reach of PINGU changes with the energy and angle resolution, as well as 
with the inclusion of the systematic uncertainties in the analysis. We will also study how 
the sensitivity to determination of the $\oct$ at PINGU goes down when one marginalizes over 
the oscillation parameters $|\ma|$, $\sch$, $\sa$ and the neutrino mass hierarchy. \footnote{Note that the 
sensitivity to the $\oct$ at PINGU is almost insensitive to the value of $\dcp$ and so we 
take $\dcpt=0$ in the simulated data for PINGU throughout this paper. For 
results that use only the PINGU data, we keep 
$\dcp=0$ fixed in the fit as well. 
In this paper we will denote the true values of the oscillation parameters 
by putting `true' within brackets.}
\\

Since the Daya Bay, RENO, 
Double Chooz and T2K experiments are already operational and since 
NO$\nu$A will also start operations soon, it is expected that we will have full or at least 
partial data from these experiments by the time we have enough data collected at 
PINGU. In addition, sizable octant sensitivity is expected from the combined results from 
T2K, NO$\nu$A and the reactor experiments as discussed above. Therefore, we will 
use the projected data from these experiments 
to give 
the sensitivity to the $\oct$ from 
these experiments combined with PINGU. 
The combination of the data from the long baseline and reactor experiments to the PINGU data 
serves four purposes. 
Firstly, we can compare and contrast the prospects of 
determining the $\oct$ at PINGU versus the prospects 
expected from the current accelerator and reactor experiments. 
Secondly, a combined analysis of PINGU with the accelerator and reactor experiments 
shows possible synergies between the different experiments. Thirdly and most importantly, the 
accelerator and reactor experiments 
severely constrain the oscillation parameters $|\ma|$, $\sin^22\theta_{23}$ and 
$\stch$. 
This reduces the effect of 
marginalization in PINGU, thereby increasing the $\chi^2$. 
Finally, the net sensitivity to the $\oct$ will anyway be given by a global analysis  
of all available 
neutrino oscillation data after a few years of running of PINGU, and  
our aim in this paper is to show this. 
\\

A discussion on the role of $\dcp$ in the analysis of the data from long baseline experiment is in order.
The sensitivity of the LBL experiments to the $\oct$ 
depends on the true value of the CP phase, $\dcpt$.
This was studied in detail in \cite{t2k2.5,animesh} 
where the octant sensitivity of 
the combined NO$\nu$A and T2K data was shown as a function of the $\dcpt$. 
In particular, in \cite{t2k2.5}, the authors showed the degeneracy between  
$\theta_{23}$ and $\dcp$, and since the value of $\dcp$ is completely unknown, the 
$\oct$ sensitivity from long baseline experiments suffers. They showed that 
running T2K in the antineutrino mode alleviates this 
$\dcp-\theta_{23}$ degeneracy issues. In particular, they compared results of 
5 years of neutrino running alone in T2K \cite{t2k}, against those obtained with 2.5 years in neutrino 
and 2.5 years in the antineutrino mode, and argued that the 2.5+2.5 years 
option for T2K was better. This issue was also recently pointed out and studied 
in \cite{machado}. 
We will also show our results for both possibilities for T2K. 
The NO$\nu$A experiment has recently re-optimized  \cite{newnova} their event selection 
criteria, following the discovery of large $\theta_{13}$ \cite{dbth13,renoth13,dcth13,t2kth13,minosth13}. 
We have used this new NO$\nu$A configuration following \cite{skanova} and 
have compared the results using the new NO$\nu$A configuration versus the 
old NO$\nu$A configuration \cite{nova}. For both NO$\nu$A configurations we use 
3+3 years of running of the experiment in the neutrino+antineutrino mode. 
A final comment is in order. 
The recent studies on the $\oct$ sensitivity from long baseline experiments 
\cite{t2k2.5,animesh}, 
imposed a prior on the value of $\theta_{13}$ in the fit and showed that the 
octant sensitivity of the long baseline experiments depends crucially on that. In 
this work we add the projected reactor data from the full 3 years run of 
Daya Bay \cite{db}, RENO \cite{reno} and Double Chooz \cite{dc}. 
\\

The paper is organized as follows. We start in section \ref{sec:pingu} with a description of 
the atmospheric neutrino events in PINGU and our method of the analysis 
of the simulated data in PINGU. In section \ref{sec:octpingu} we present the 
octant sensitivity from PINGU alone, allowing for full marginalization of the oscillation 
parameters. We discuss in detail the impact of detector resolutions, systematic uncertainties 
and uncertainties in oscillation parameters on the final octant sensitivity at PINGU. 
In section \ref{sec:octcombined} we begin with a discussion of the octant sensitivity 
of T2K+NO$\nu$A+Reactor experiments. We compare and contrast the reach of old NO$\nu$ 
versus new NO$\nu$A configuration. We also make a comparative study of the octant sensitivity 
with 5 years of T2K running in the neutrino mode alone vis-a-vis 2.5 years of T2K running in neutrino 
and remaining 2.5 years of T2K running in the antineutrino mode. 
Finally, we show the $\oct$ sensitivity coming from combined fits of the PINGU data along with 
the projected runs of the T2K+NO$\nu$A+Reactor experiments. We end in 
section \ref{sec:concl} with our conclusions.

\section{\label{sec:pingu}Atmospheric Neutrinos in PINGU}

The number of atmospheric 
muon events expected in PINGU after $T$ years of running of the experiment 
is given as
\be
(N_\mu)_{ij}&=&2\pi \,N_T \,T\int_{(\cos\Theta)_i}^{(\cos\Theta)_{i+1}}d\cos\Theta
\int_{E_j}^{E_{j+1}}dE \,\epsilon \,\rho_{ice}\,V_{eff}(E) \times\nn\\
&&\bigg[\left(\frac{d^2\phi_\numu}{d\cos\Theta'dE'}P_{\mu\mu} + 
\frac{d^2\phi_\nue}{d\cos\Theta'dE'}P_{e\mu} \right)\,\sigma_{CC}(E) + \nn\\
&&\bigg[\left(\frac{d^2\phi_\anumu}{d\cos\Theta'dE'}P_{\bar\mu\bar\mu} + 
\frac{d^2\phi_\anue}{d\cos\Theta'dE'}P_{\bar{e}\bar\mu} \right)\bigg]\,\sigma_{CC}(E)
\,
\label{eq:events}
\ee
where $N_T$ are the number of targets in the detector, $T$ is the exposure time, 
$\epsilon$ is the detector efficiency, $\rho_{ice}V_{eff}$ is
the effective mass of the detector, $V_{eff}$ is the effective volume, 
$d^2\phi_\alpha/dE'd\cos\Theta'$ is the neutrino flux of flavor $\alpha$ \cite{honda}, 
$\sigma_{CC}(E)$ is the neutrino-nucleon interaction cross-section, 
$R(E,E')$ and $R(\Theta,\Theta')$ are 
the energy and angle resolution functions of the detector respectively, 
and $P_{\mu\mu}$ and $P_{e\mu}$ are 
the muon neutrino survival probability and electron neutrino 
to muon neutrino conversion probabilities, respectively. 
The resolution functions relate the true energy $E'$ and zenith angle $\Theta'$ 
with the reconstructed 
energy $E$ and zenith angle $\Theta$, of the neutrino. We assume Gaussian functional 
form for the resolutions 
functions with widths 
\be
\sigma_E = a+bE'\,,~~~~\sigma_\Theta = c\cdot\sqrt{\frac{1~{\rm GeV}}{E'}}\,.
\label{eq:resol}
\ee
Since the final resolution widths for PINGU is still being estimated from detailed simulations, 
we will assume two sets of values for the energy resolution width $\sigma_E$ corresponding to 
$a=0,~b=0.2$ and $a=2,~b=0$, and two sets of values for the angle resolution width $\sigma_\Theta$ 
corresponding to $c=1$ and $c=0$, where $c$ (and hence $\sigma_\Theta$) is in radians. 
These values for the resolution functions agree with that 
used in the literature \cite{pingumh,ww,bsmh}.  The effective mass of the detector is read 
from \cite{pingutalk} for the curve labelled ``Triggered Effective Volume, R=100m". 
The probabilities $P_{\mu\mu}$ and 
$P_{e\mu}$ are calculated numerically solving the propagation equation of the neutrinos 
through the atmosphere and inside the earth, and using the PREM profile \cite{prem} 
for the earth matter density. For simplicity we take $\epsilon=1$, since any flat $\epsilon$ can 
be easily adjusted against the exposure taken at the detector. 
The index $i$ in Eq. (\ref{eq:events}) runs 
over the number of $\cos\Theta$ bins in the data while $j$ runs over 
the number of energy bins. In the analysis presented in this paper, we have 
taken 10 equal zenith angle bins in $\cos\Theta$ between $\cos\Theta=-1$ and $0$, 
and 7 equal energy bins between $E=2$ and $14$ GeV. 
The data is generated for the oscillation parameters 
given in Table \ref{tab:param} for either the normal or inverted hierarchy and for a 
given value of $\sa$. Note that we use $\meff >0$ as our definition of the normal 
mass hierarchy and $\meff<0$ as inverted mass hierarchy, where 
\be
\meff= \ma - (\cos^2\theta_{12} - \cos\delta_{CP}\sin\theta_{13}\sin2\theta_{12}\tan\theta_{23})\ms
\,,
\label{eq:meff}
\ee
The simulated data is then fitted with the wrong octant solution of $\sa$, 
allowing the test $|\meff|$, $\sch$, $\sa$ as well as the neutrino mass hierarchy 
to vary in the fit. The statistical fit is 
performed using a $\chi^2$ function defined as 
\be
\chi^2={\min_{\{\xi_j\}}} \displaystyle\sum\limits_{ij} \left
[\frac{\left(N_{ij}^{\prime{th}}-N_{ij}^{ex}\right) ^2}{N_{ij}^{ex}}
\right] + \displaystyle\sum\limits_{s=1}^k \xi_s^{2}
\,,
\label{eq:chisq}
\ee
\be
N_{ij}^{\prime{th}}=N_{ij}\left(1+\displaystyle\sum\limits_{s=1}^k \pi_{ij}^s{\xi_s}\right) + 
{\cal O}(\xi_s^2)
\,,
\label{eq:evth}
\ee
where $N_{ij}^{ex}$ 
is the observed number of muon events in the $i^{th}$ $\cos\Theta$ and 
$j^{th}$ energy bin 
and 
$N_{ij}^{{th}}$ is the 
corresponding theoretically predicted event 
spectrum for the wrong octant solution of $\theta_{23}$. The 
 $\pi^s_{ij}$ in Eq. (\ref{eq:evth}) 
 is the ${ij}^{th}$ systematic uncertainty in the $i^{th}$ $\cos\Theta$ and 
 $j^{th}$ energy bin and 
$\xi_s$ is the pull variable corresponding to the uncertainty $\pi^s_{ij}$. 
We have included five systematic uncertainties in our analysis. They are, an overall 
flux normalization error of 20\%,  a cross-section uncertainty of 10\%, a 5\% uncertainty on 
the zenith angle dependence of the fluxes,  
an energy dependent ``tilt factor", and a 5\% additional overall uncertainty.
The parameters $|\meff|$ and $\sch$ are 
varied freely in the fit in the 
range given in Table \ref{tab:param}. For $\sa$, we marginalize around $\Delta\sa \pm 0.1$ 
around the wrong octant $\sa$ for any given $\sa$(true). Finally, the $\chi^2$ is 
computed for both the test hierarchies and the minimum $\chi^2$ chosen. 

\begin{table}[h]
\begin{center}
\begin{tabular}{l|c|c}
\hline
&\\[-0.9mm]
Parameter & True value used in data & $3\sigma$ range used in fit \\[2mm] \hline
&&\\
[-0.9mm]$\ms$ & $7.5 \times 10^{-5}$ eV$^2$ & $[7.0-8.0] \times 10^{-5}$ eV$^2$\\[2mm]
$\sin^2\theta_{12}$ & 0.3 & $[0.265-0.33]$\\[2mm]
$|\meff|$ & $2.4 \times 10^{-3}$ eV$^2$ & $[2.1-2.6]\times 10^{-3}$ eV$^2$\\[2mm]
$\delta_{CP}$ & 0 & $[0 - 2\pi]$\\[2mm]
$\stch$ & 0.1 & $[0.07-0.13]$   \\[2mm]
\hline
\end{tabular}
\caption{\label{tab:param}
Benchmark true values of oscillation parameters 
used in the simulations, unless otherwise stated. The range over which they are 
allowed to vary freely in the fit is also shown in the last column. For $\sat$ 
we marginalize around $\Delta\sa \pm 0.1$ 
around the wrong octant $\sa$ for any given $\sa$(true). For PINGU alone analysis 
we keep $\ms$, $\sss$ and $\dcp$ fixed at their true values. 
}
\end{center}
\end{table}

\section{\label{sec:octpingu}Octant Sensitivity from PINGU Alone}

\begin{figure}
\centering
\includegraphics[width=0.495\textwidth]{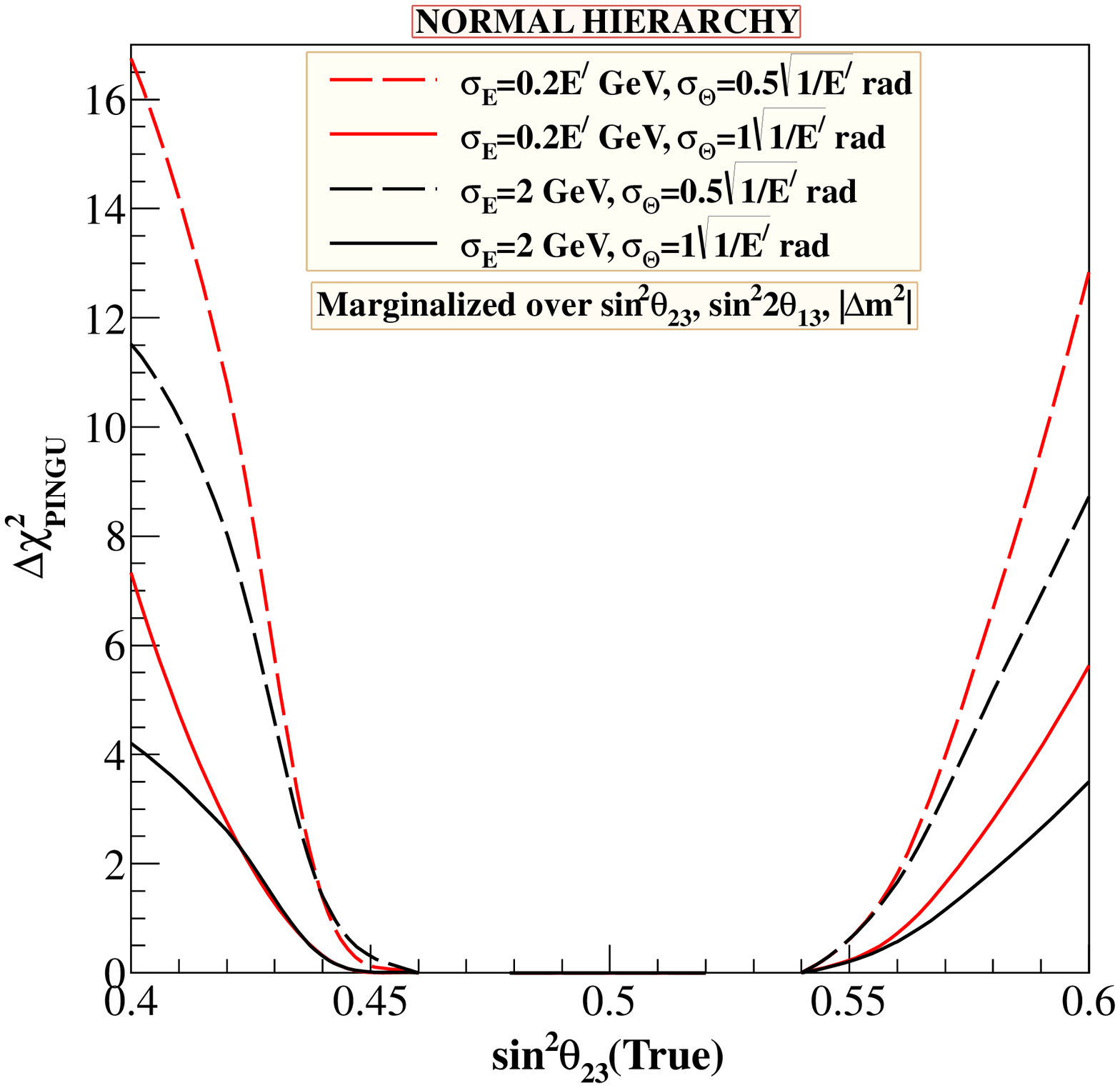}
\includegraphics[width=0.495\textwidth]{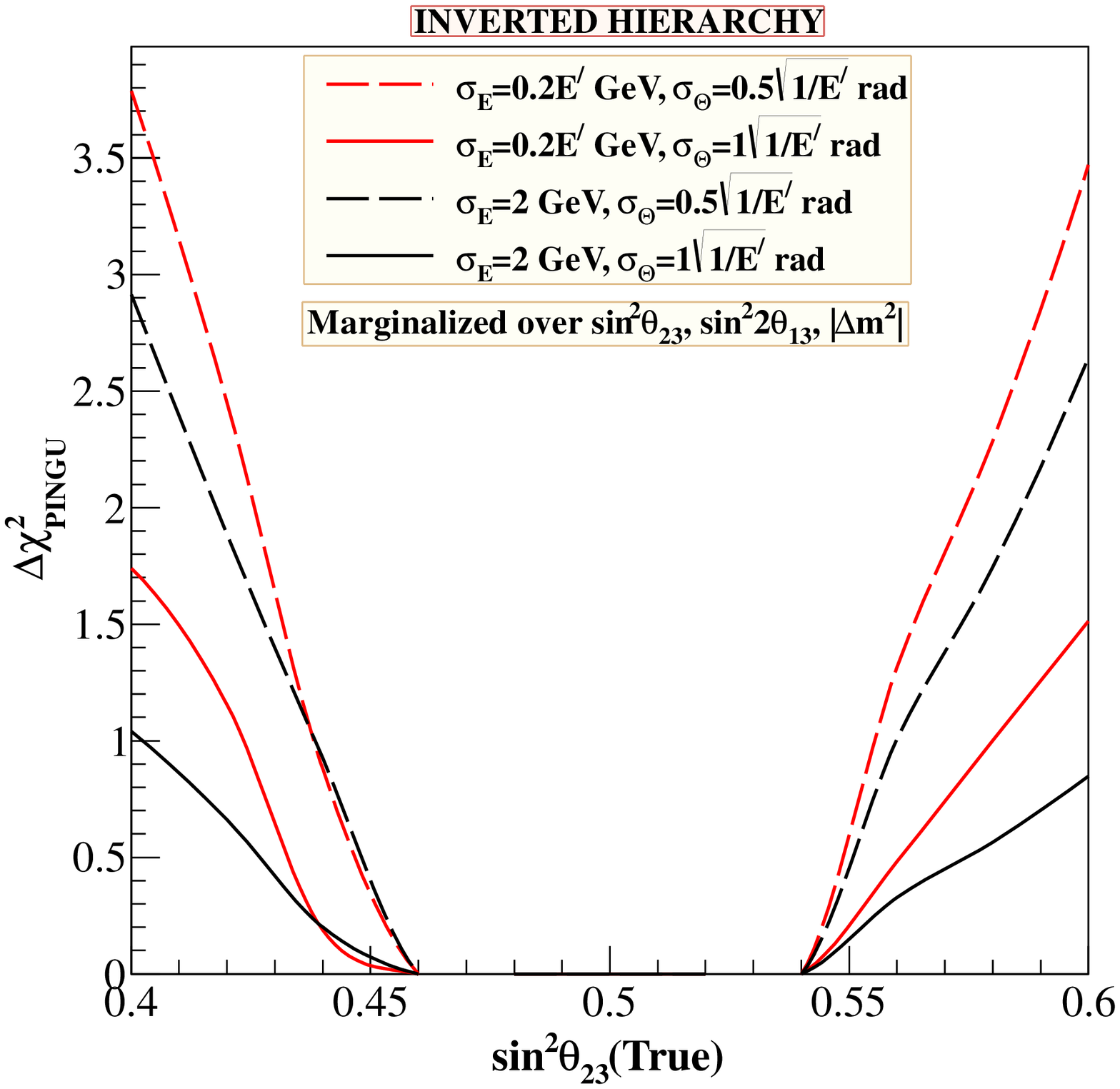}
\includegraphics[width=0.495\textwidth]{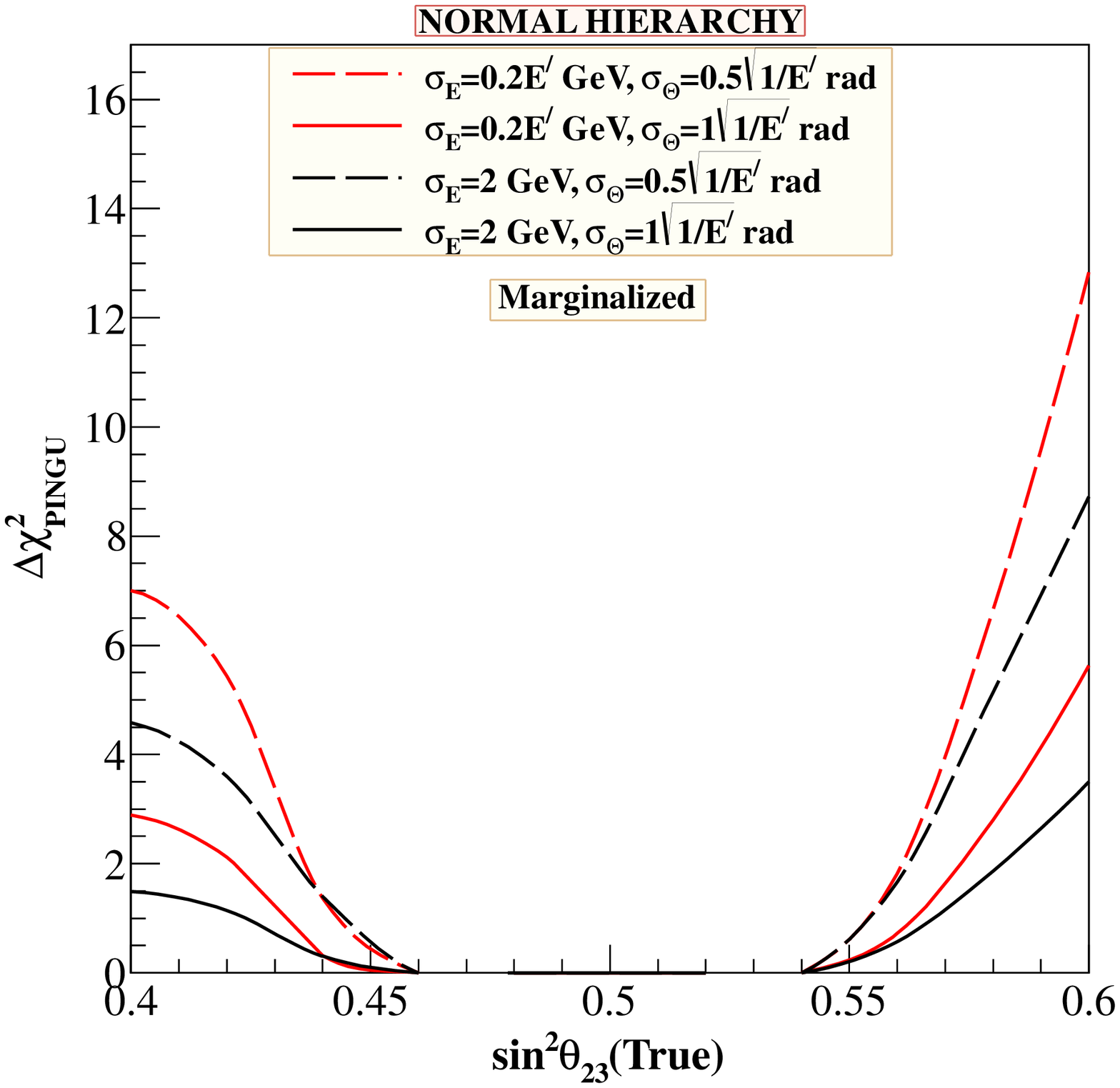}
\includegraphics[width=0.495\textwidth]{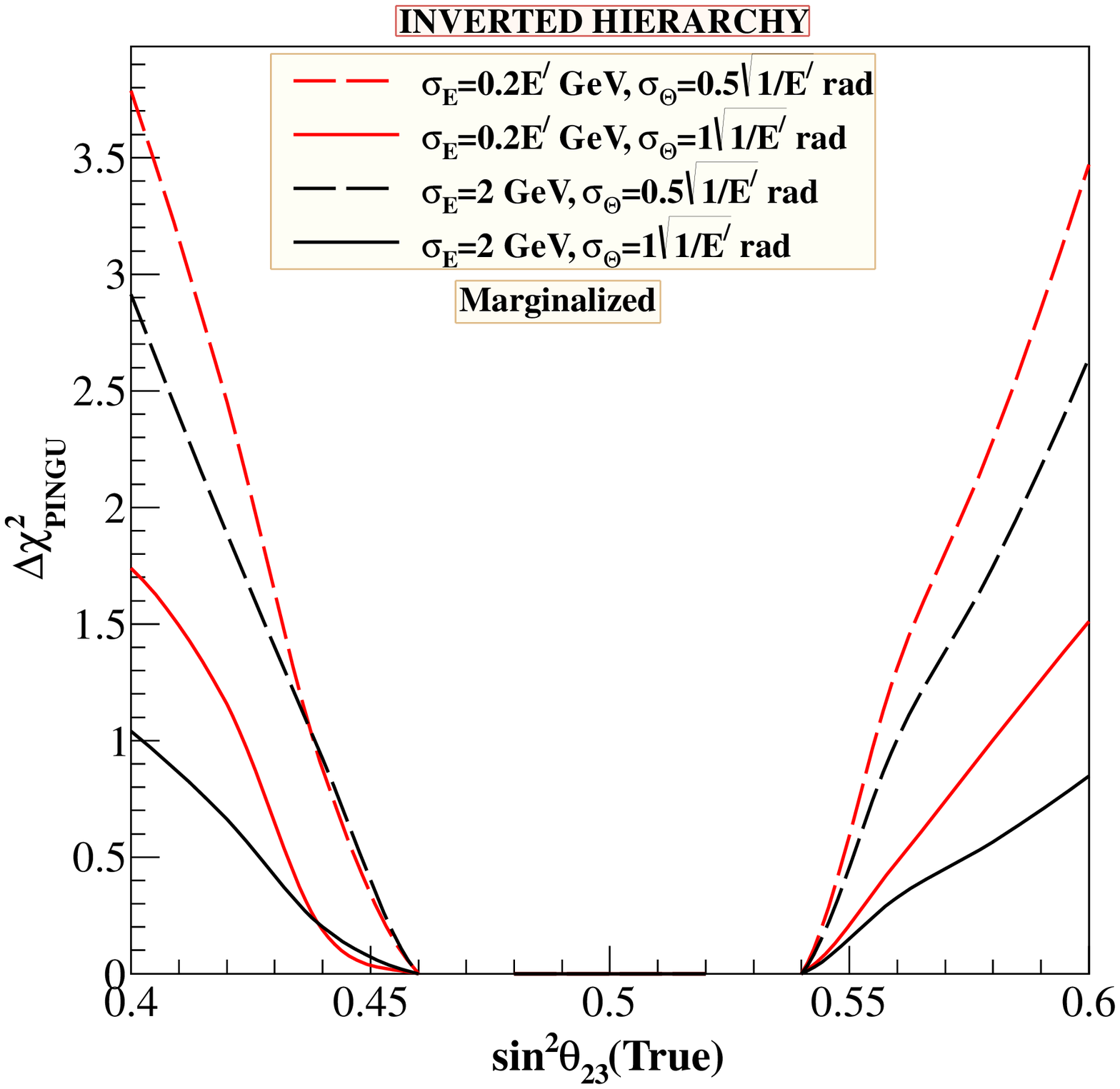}
\caption{\label{fig:senspingu}
$\Delta \chi^2$ for the wrong octant obtained from just the 
PINGU data with 3 years statistics, as a function of $\sat$. The left panel is 
for normal hierarchy taken as true while the right panel is for inverted hierarchy taken 
as true. In the top panels, the $\chi^2$ is marginalized over the oscillation parameters, $|\meff|$, $\stch$ and 
$\sa$ as described in the text, but the mass hierarchy is held fixed to the assumed true case in the fit. 
In the bottom panels, the $\chi^2$ is marginalized over the oscillation parameters, $|\meff|$, $\stch$ and 
$\sa$, as well as the mass hierarchy, keeping mass hierarchy free in the fit.
The four lines are for the four possible combinations for the 
choices of the energy and angle resolution of PINGU. }
\end{figure}

The results of our statistical analysis and the sensitivity of PINGU {\it alone} to the octant of 
$\theta_{23}$ is given in Fig. \ref{fig:senspingu}, where we have shown the 
$\Delta \chi^2$ as a function of $\sat$. We have used 3 years of running of the experiment 
with the effective mass of the detector read 
from \cite{pingutalk} for the curve labelled ``Triggered Effective Volume, R=100m". 
The left panels of this figure give the octant sensitivity when the data is generated assuming 
the normal hierarchy ($\meff>0$), while the right panels are obtained when the data is 
generated assuming the inverted mass hierarchy ($\meff<0$). We have included the 
full flux systematic uncertainties as outlined in the previous section. The top panels are 
obtained when  
the $\chi^2$ is marginalized over the oscillation parameters, $|\meff|$, $\stch$ and 
$\sa$ as described in the previous section, however, the test mass hierarchy is kept fixed at the 
assumed true value. The bottom panels are obtained when in addition to $|\meff|$, $\stch$ and 
$\sa$, the $\Delta \chi^2$ is marginalized also with respect to the mass hierarchy. 
The four lines are for the four possible combinations for  
the energy and angle resolutions of PINGU. The red dashed lines are for {\it our}  
optimal choice of $\sigma_E = 0.2E'$ and $\sigma_\Theta = 0.5\,(1~{\rm GeV}/\sqrt{E'})$, as 
discussed in section \ref{sec:pingu}. 
\\

This figure shows the impact of a variety of things on the octant sensitivity at PINGU. 
Let us start by looking at the impact of the resolution functions on the octant sensitivity. 
From the top left panel, we find that if the normal hierarchy was true and the test hierarchy 
was kept fixed as normal, then 
the wrong $\theta_{23}$ octant could be ruled out at more than $3\sigma$ C.L. \footnote{Where 
the number of $\sigma$ $n$ is defined as $n\sigma = \sqrt{\Delta \chi^2}$.}
if $\sat < 0.424$ and $\sat > 0.588$ if 
$\sigma_E = 0.2E'$ and $\sigma_\Theta = 0.5\,(1~{\rm GeV}/\sqrt{E'})$. 
To show the impact of the resolution functions on the 
octant sensitivity, we have repeated the exercise for different combinations of $\sigma_E$ and 
$\sigma_\Theta$. Keeping $\sigma_\Theta=0.5/\sqrt{E'}$ and changing $\sigma_E$ from 
$0.2E'$ to $2$ GeV reduces the statistical significance of the octant measurement 
and now we expect $3\sigma$ sensitivity for $\sat < 0.416$ and $\sat > 0.6$. 
The impact of worsening of the zenith angle sensitivity is seen to be sharper. 
Keeping $\sigma_E=0.2E'$ if we change $\sigma_\Theta=0.5\sqrt{1~GeV/E'}$ to 
$\sigma_\Theta=1.0\sqrt{1~GeV/E'}$, this 
reduces the statistical significance of the octant measurement significantly 
and now we have only $2\sigma$ sensitivity for $\sat < 0.412$ and $\sat > 0.588$. 
For the case where both energy resolution and angle resolutions are 
worsened to $\sigma_E=2$ GeV and $\sigma_\Theta = 1.0\sqrt{1~GeV/E'}$, we find that 
the wrong octant can be ruled out only at $2\sigma$ for $\sat <0.4$ and $\sat > 0.6$. 
\\

The right panels of Fig. \ref{fig:senspingu} show the sensitivity of PINGU alone 
to the $\oct$ if the inverted hierarchy was true. We find that 
for the inverted hierarchy the sensitivity falls significantly, 
and even for $\sat=0.4$, the wrong hierarchy would be barely ruled at the $2\sigma$ C.L.. 
The sharp difference between the reach of the experiment to the $\oct$ between the 
normal and inverted hierarchy can be understood as follows. 
Let us  
say that we generate the data at a certain value of $\theta_{23}$ and 
fit the data for for $\theta_{23}(fit)$. 
Since the main sensitivity of PINGU to the $\oct$ 
comes from earth matter effects, for normal hierarchy the difference between number of 
events for $\theta_{23}$ and $\theta_{23}(fit)$ is expected to be non-zero for the 
neutrino events and nearly zero for the antineutrino events. 
In that case, since PINGU cannot distinguish between the particle and antiparticle, 
the $\chi^2$ obtained for normal hierarchy can be very roughly written as
\be
\chi^2(NH) \sim \frac{(N_{\mu^-}^{NH}(\theta_{23}) - N_{\mu^-}^{NH}(\theta_{23}^{fit}))^2}{N_{\mu^-}^{NH}(\theta_{23}) - 
N_{\mu^+}^{NH}(\theta_{23})}
\,,
\label{eq:nh}
\ee
where for simplicity, 
we have neglected the systematic uncertainties and have assumed the test hierarchy to be fixed. 
The same argument holds in the general case.
In Eq. \ref{eq:nh} we have taken 
$N_{\mu^+}^{NH}(\theta_{23})-N_{\mu^+}^{NH}(\theta_{23}^{fit}) \simeq 0$, since for normal hierarchy 
there are almost no earth matter effects in the antineutrino channel. 
On the other hand for the inverted hierarchy the corresponding $\chi^2$ is approximately given as
\be
\chi^2(IH) \sim \frac{(N_{\mu^+}^{IH}(\theta_{23}) - N_{\mu^+}^{IH}(\theta_{23}^{fit}))^2}{N_{\mu^-}^{IH}(\theta_{23}) - N_{\mu^+}^{IH}(\theta_{23})}
\,,
\label{eq:ih}
\ee
since in the case of the inverted hierarchy the difference between the number of events 
$N_{\mu^-}^{IH}(\theta_{23})-N_{\mu^-}^{IH}(\theta_{23}^{fit}) \simeq 0$, as for this case 
there are almost no earth matter effects in the neutrino channel.
We can very approximately write
\be
\bigg(N_{\mu^-}^{NH}(\theta_{23}) - N_{\mu^-}^{NH}(\theta_{23}^{fit})\bigg) \simeq \Delta P_{\mu\mu}^{NH}N_{\mu^-}^0 + 
\Delta P_{e\mu}^{NH}N_{e^-}^0
\,,
\label{eq:nhdiff}
\ee
where $N_{\mu^-}^0$ and 
 $N_{e^-}^0$ are the number of $\numu$-events and $\nue$-events in absence of oscillations and 
\be
\Delta P_{\mu\mu}^{NH} &=& P_{\mu\mu}^{NH}(\theta_{23}) - P_{\mu\mu}^{NH}(\theta_{23}(fit))\,,\nn \\  
\Delta P_{e\mu}^{NH} &=& P_{e\mu}^{NH}(\theta_{23}) - P_{e\mu}^{NH}(\theta_{23}(fit))\,.
\ee
For the inverted hierarchy case we have
\be
(N_{\mu^+}^{IH}(\theta_{23}) - N_{\mu^+}^{IH}(\theta_{23}^{fit})) \simeq N_{\mu^+}^0\,\Delta P_{\bar\mu\bar\mu}^{IH} + 
N_{e^+}^0\,\Delta P_{\bar{e}\bar\mu}^{IH}
\,,
\label{eq:ihdiff}
\ee
where $N_{\mu^+}^0$ and 
 $N_{e^+}^0$ are the number of $\anumu$-events and $\anue$-events in absence of oscillations and 
$\Delta P_{\bar\mu\bar\mu}^{IH}$ and $\Delta P_{\bar{e}\bar\mu}^{IH}$ are given by 
\be
\Delta P_{\bar\mu\bar\mu}^{IH} &=& P_{\bar\mu\bar\mu}^{IH}(\theta_{23}) - 
P_{\bar\mu\bar\mu}^{IH}(\theta_{23}(fit))\,,\nn \\  
\Delta P_{\bar{e}\bar\mu}^{IH} &=& P_{\bar{e}\bar\mu}^{IH}(\theta_{23}) - 
P_{\bar{e}\bar\mu}^{IH}(\theta_{23}(fit))\,.
\ee
Since 
\be
\Delta P_{\mu\mu}^{NH} \simeq \Delta P_{\bar\mu\bar\mu}^{IH}~~{\rm and}~~
\Delta P_{e\mu}^{NH} \simeq \Delta P_{\bar{e}\bar\mu}^{IH}
\ee
and
\be
N_{\mu^+}^0 \simeq \frac{1}{2}N_{\mu^-}^0~~{\rm and}~~
N_{e^+}^0 \simeq \frac{1}{2}N_{e^-}^0
\,
\ee
we have from Eqs. (\ref{eq:nh}) and (\ref{eq:ih})
\be
\chi^2(IH) \sim \frac{1}{4}\cdot \chi^2(NH)\cdot \frac{N_{\mu^-}^{NH}(\theta_{23}) - N_{\mu^+}^{NH}(\theta_{23})}
{N_{\mu^-}^{IH}(\theta_{23}) - N_{\mu^+}^{IH}(\theta_{23})}
\,.
\label{eq:chicompare}
\ee
Since the number of neutrino  ($\mu^-$) events are more than twice the number of 
antineutrino ($\mu^+$) events, and since due to earth matter effects 
$N_{\mu^-}^{NH} < N_{\mu^-}^{IH}$ for any value of $\theta_{23}$, 
\be
\frac{N_{\mu^-}^{NH}(\theta_{23}) - N_{\mu^+}^{NH}(\theta_{23})}
{N_{\mu^-}^{IH}(\theta_{23}) - N_{\mu^+}^{IH}(\theta_{23})} < 1
\ee
and from Eq. (\ref{eq:chicompare}) we see that we expect 
\be
\chi^2(IH) < \frac{ \chi^2(NH)}{4}
\,.
\ee
This rough comparison between the expected octant sensitivity between normal and inverted 
mass hierarchy cases is seen to agree rather well with the results shown in 
Fig. \ref{fig:senspingu}, which have been obtained from a detailed numerical computation. 
\\

Finally, a comparison of the upper panels with the lower panels of Fig. \ref{fig:senspingu} 
shows the impact of marginalization over the test mass hierarchy in the fit. 
Let us first discuss the case where the data is generated for true inverted hierarchy.
For the true inverted hierarchy case we can see from the figure that marginalization 
over the test mass hierarchy in the  fit has no impact on the octant sensitivity of 
the experiment. 
However,  
for true normal hierarchy there is significant reduction in the $\Delta \chi^2$ for  
low values of $\sat$. In particular, we can see that for $\sat=0.4$ the statistical 
significance of the octant determination from 3 years of PINGU data alone,
comes down from $\Delta \chi^2=16.75$ to $\Delta \chi^2=6.9$ for the optimal 
resolution case of  $\sigma_E = 0.2E'$ and $\sigma_\Theta = 0.5\,(1~{\rm GeV}/\sqrt{E'})$. 
For the other choices of the combination for 
$\sigma_E$ and $\sigma_\Theta$ also we see a similar trend, wherein the data with 
true normal hierarchy is fitted with the wrong test inverted hierarchy and with a lower 
$\chi^2$, reducing thereby the octant sensitivity from PINGU alone. 
However, for $\sat > 0.44$ the marginalization over hierarchy does not have any 
impact what-so-ever on the octant sensitivity of PINGU, even for the case of 
true normal hierarchy. 
\\
 
\begin{figure}
\centering
\includegraphics[width=0.495\textwidth]{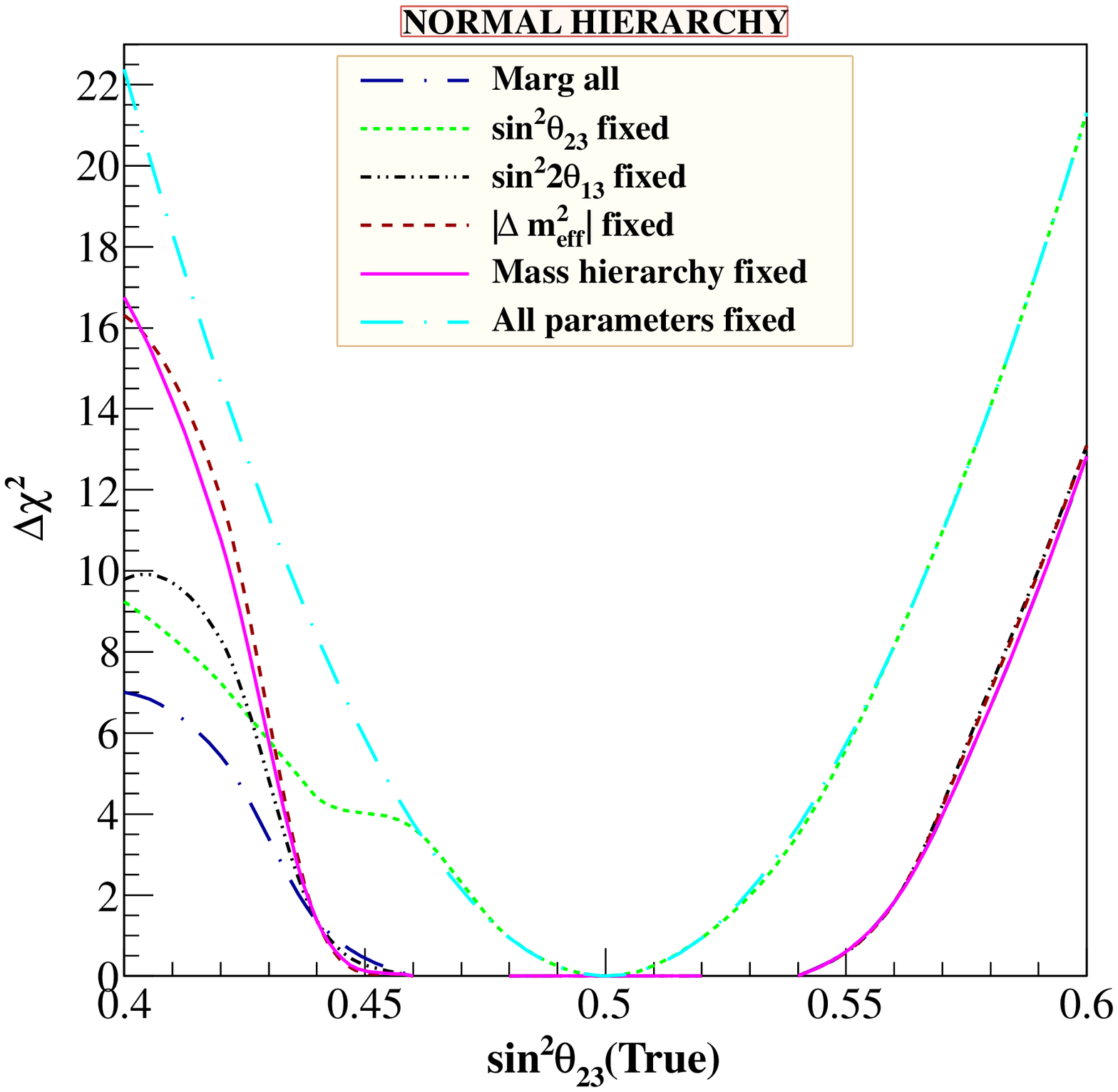}
\includegraphics[width=0.495\textwidth]{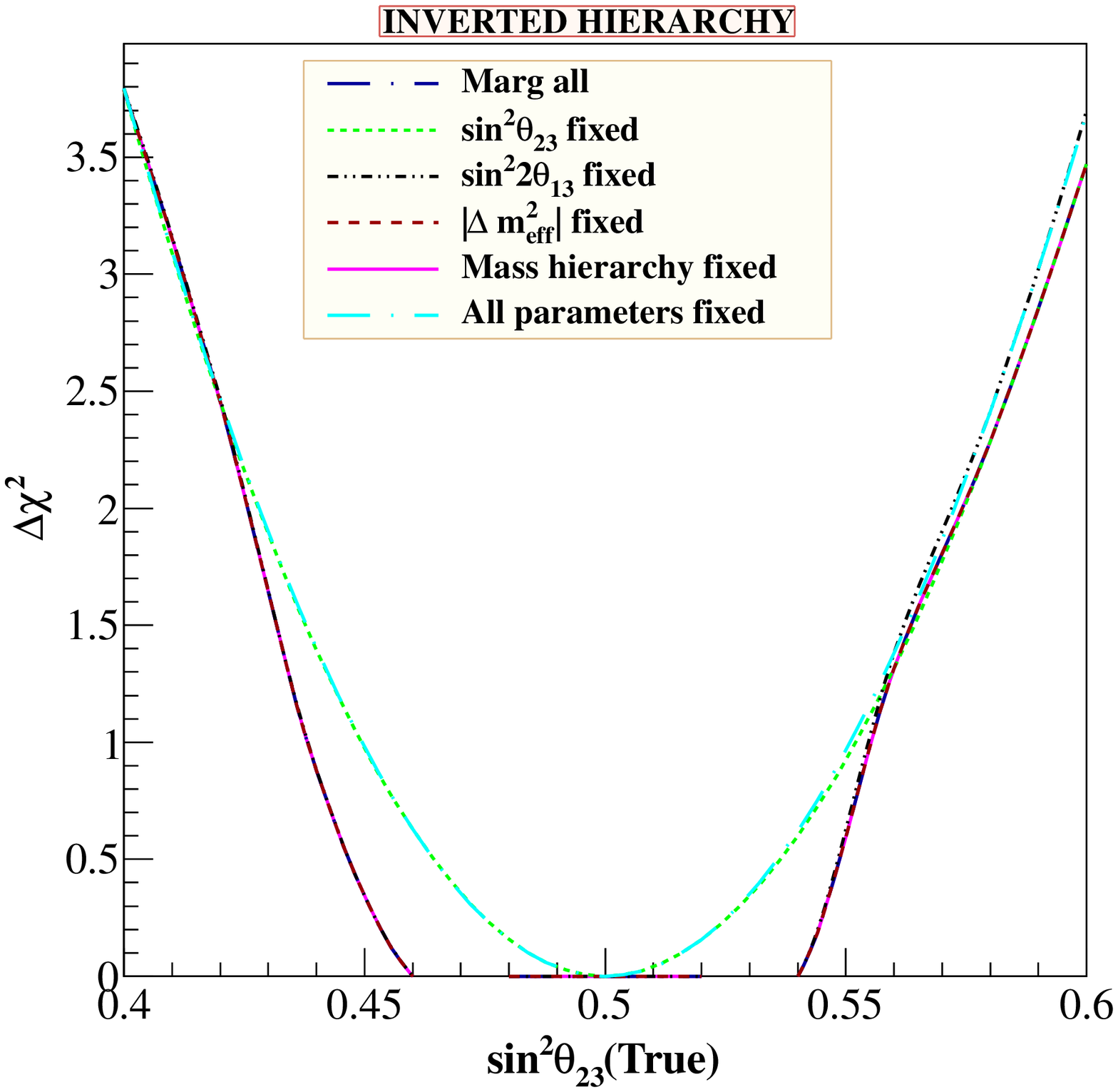}
\caption{\label{fig:senspingu_marg}
Impact of marginalization over the different oscillation parameters on the 
$\Delta \chi^2$ for the wrong octant obtained from just the 
PINGU data with 3 years statistics, as a function of $\sat$. The left-panel is 
for data generated with normal hierarchy while the right-panel is for data generated with 
inverted hierarchy.}
\end{figure}

Note from Fig. \ref{fig:senspingu} that the $\Delta\chi^2$ is large for $\sat$ significantly 
far from maximal mixing. However, the statistical significance falls very fast and 
nearly disappears for $\sat \gtap 0.45$ and $\sat \ltap 0.55$. The main reason for this 
flat $\Delta \chi^2 \simeq 0$ between these two limiting regions of $\sat$ is the marginalization 
over the mixing angle $\theta_{23}$ in the fit. We show the effect of marginalization on the 
$\oct$ sensitivity in Fig. \ref{fig:senspingu_marg}. We have shown this figure for 
$\sigma_E = 0.2E'$ and $\sigma_\Theta = 0.5\,(1~{\rm GeV}/\sqrt{E'})$. 
The left-hand panel of this figure is for true normal hierarchy while the right-hand 
panel is for true inverted hierarchy case.  
The blue long dot-dashed  
lines in the two panels of this figure 
are the same as the red dashed line in the bottom panels of Fig. \ref{fig:senspingu}, 
where the $\chi^2$ is 
marginalized over all oscillation parameters in the range given in Table \ref{tab:param}, as well 
as over the test neutrino mass hierarchy.  
The cyan long dot-dashed solid lines are 
obtained when all oscillation parameters are taken fixed in the fit. We find that 
in that case the sensitivity shows a remarkable improvement, especially for the 
true normal hierarchy case. For $\sat=0.4$ the $\Delta \chi^2$ jumps from 
about 7 for the fully marginalized case to more than 22 for the all parameter fixed
case. For the true inverted hierarchy case, the difference between the fully marginalized 
line and the all parameters fixed line is less dramatic, with the difference coming only 
for values of $\sat$ close to maximal mixing. 
The other lines in this figure have been 
drawn by keeping only one oscillation parameter fixed at a time, leaving the others to vary. 
For the true inverted hierarchy case, 
maximum difference to the plots, qualitatively as well as quantitatively, 
comes on fixing $\theta_{23}$ in the fit, which is shown by the green dashed lines. 
Fixing of this parameter increases the 
octant sensitivity considerably in terms of the range of $\sat$ for which 
the wrong octant solution can be disfavored, 
and we get a sensitivity which is very close to what 
we would have obtained if all oscillation parameters were fixed. 
In particular, the $\Delta\chi^2$ now has a softer fall as 
we approach maximal mixing angle. For the true inverted hierarchy 
case, the right-hand panel reveals that fixing any of the other 
oscillation parameter or the neutrino mass hierarchy makes no 
impact on the $\Delta \chi^2$. 
\\

For the true normal hierarchy case, 
the effect of fixing one parameter at a time is seen to have a much 
richer behavior, especially for low $\sat$. This is because the 
effect of marginalization over the oscillation parameters as well the 
mass hierarchy is much larger for the true normal hierarchy case. 
In particular, for low values of $\sat \ltap 0.45$, we find that 
fixing the neutrino mass hierarchy (pink solid line) causes a significant increase 
in the $\Delta \chi^2$, as we had seen before in Fig. \ref{fig:senspingu}. 
Fixing $|\meff|$ (red short dashed line) is also seen to give a similar change to the 
$\Delta \chi^2$. The reason for this is as follows. When we let the test 
hierarchy vary in the fit, the wrong inverted hierarchy is preferred by pushing 
the fit value of $\theta_{13}$ to its current allowed upper limit as given in Table \ref{tab:param}, 
and by suitably adjusting the fit $|\meff|$. When we fix $|\meff|$, then this freedom 
is lost, and in that case even if we allow the test hierarchy to vary, the minima 
still comes for the test normal hierarchy. Therefore, the fixed $|\meff|$ case looks 
very similar to the fixed hierarchy case. The other parameter which has a 
significant impact for low values of $\sat$ is $\theta_{13}$. The black dot-dashed lines 
show that the $\Delta \chi^2$ increases when $\theta_{13}$ is fixed at its true 
value in the fit. This point is worth noting. The reason is that constraint on 
$\theta_{13}$ is expected to significantly improve with more data from the reactor 
experiments, particularly Daya Bay, as well as T2K and NO$\nu$A. In the next 
section when we will add the reactor data to the PINGU data we will find a 
sensitivity curve very close to the one shown in this figure for fixed $\theta_{13}$. 
Finally, as in the case of the true inverted hierarchy, fixing $\theta_{23}$ in the 
fit brings a large effect on the octant sensitivity of PINGU. In fact, for the high 
octant region of the curve, we see that only the $\theta_{23}$ parameter 
plays a role in the marginalization process. 
\\

\begin{figure}
\centering
\includegraphics[width=0.495\textwidth]{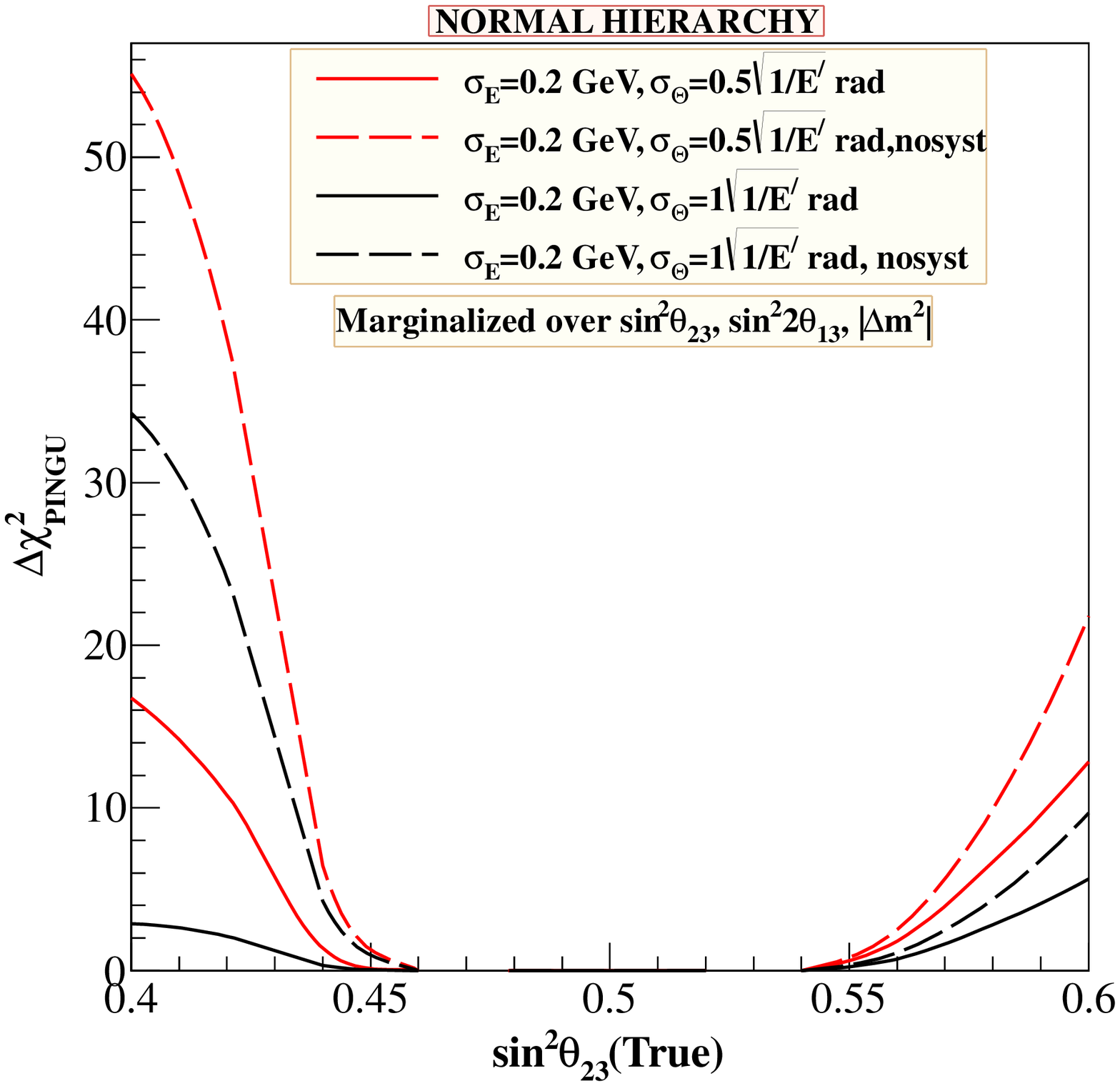}
\includegraphics[width=0.495\textwidth]{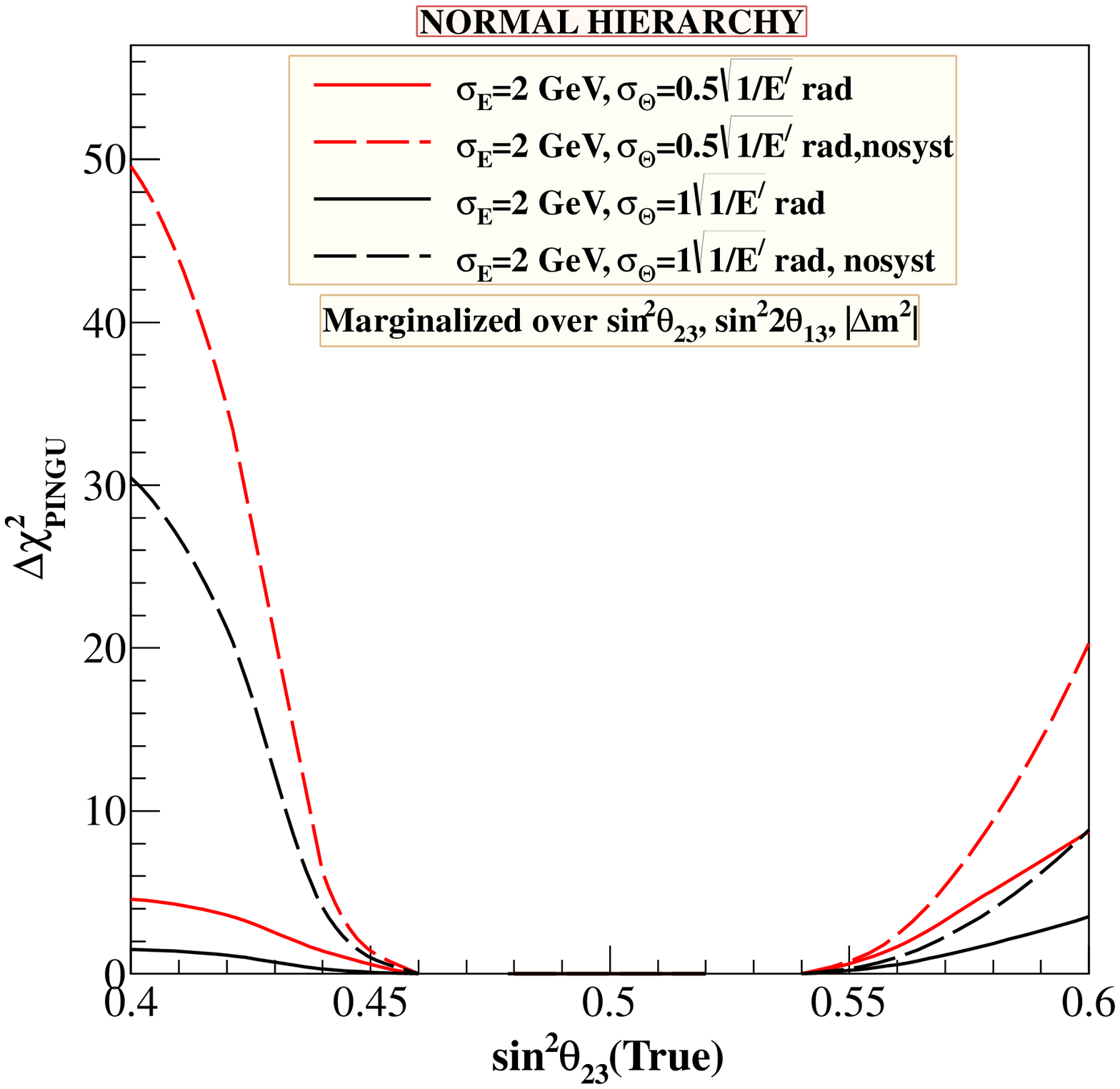}
\includegraphics[width=0.495\textwidth]{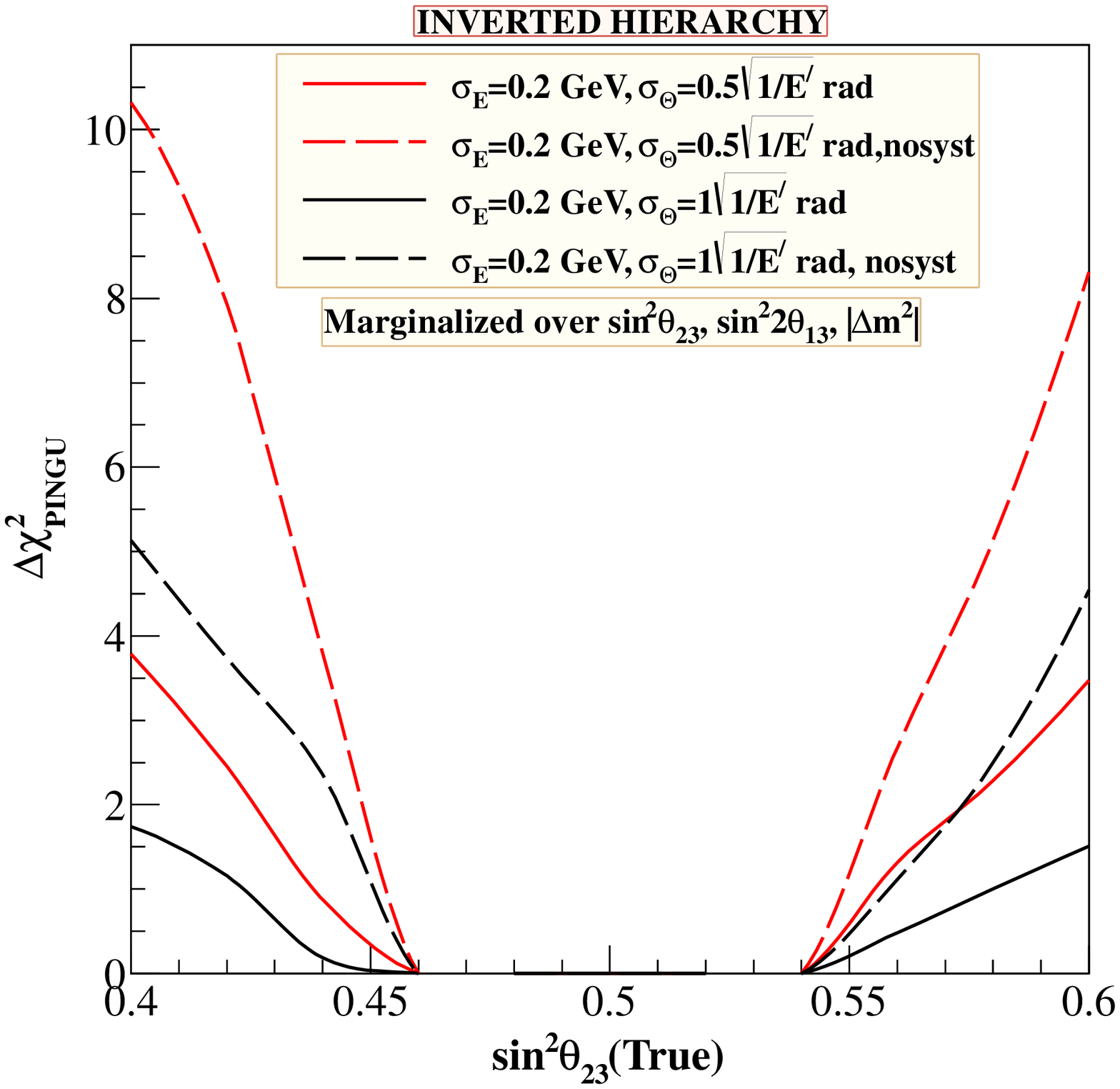}
\includegraphics[width=0.495\textwidth]{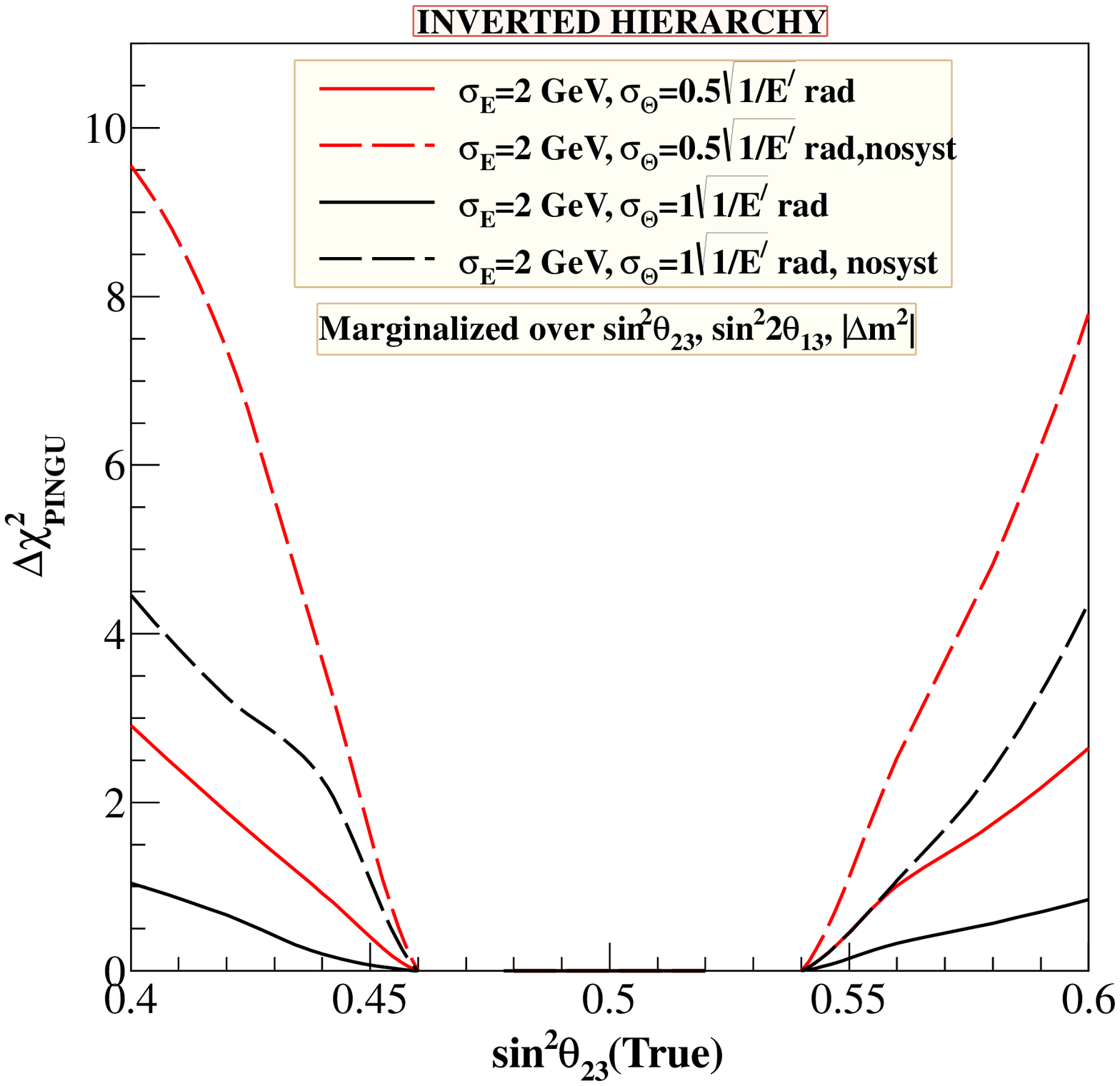}
\caption{\label{fig:pingusyst}
Impact of systematic uncertainties on the 
$\Delta \chi^2$ for the wrong octant obtained from just the 
PINGU data with 3 years statistics, as a function of $\sat$. The solid lines 
show the $\Delta \chi^2$ when systematic uncertainties are included, while the 
dashed lines are obtained when we do not include the systematic uncertainties. The 
assumption of the width of the resolution functions are shown in the figure legend. We 
show the results for all combinations of $\sigma_E$ and $\sigma_\Theta$. The upper panels 
are for normal hierarchy while the lower panels are for inverted hierarchy. The test hierarchy 
is kept fixed at the true value for all cases.}
\end{figure}

In order to show the impact of the systematic uncertainties on the $\oct$ sensitivity at 
PINGU, we plot in Fig. \ref{fig:pingusyst} the $\Delta \chi^2$ as a function of $\sat$, with (solid lines)
and without (dashed lines) systematic uncertainties. For simplicity we show the 
plot for the case where the mass hierarchy is kept fixed at the true value in the fit. 
The widths of the resolution functions 
assumed for each case is shown in the figure legend. 
The upper panels in this figure are for the 
normal hierarchy case while the lower panels show the result for the true inverted hierarchy. 
We can see that for the normal hierarchy case, systematic uncertainties reduce the 
$\Delta \chi^2$ by a factor of about 3.5 when $\sigma_E = 0.2E'$ GeV (upper left-hand panel) 
and for both $\sigma_\Theta = 0.5\sqrt{1/E'}$ and $\sigma_\Theta = 1\sqrt{1/E'}$. 
For the worse energy resolution case of $\sigma_E = 2$ GeV (upper right-hand panel) 
the systematic uncertainties reduce the $\Delta \chi^2$ by a factor of about 4.2
for $\sigma_\Theta = 0.5\sqrt{1/E'}$ and by a factor of about 7.5 for 
$\sigma_\Theta = 1\sqrt{1/E'}$. Notice that the impact of the systematic uncertainty 
increases as the energy and angle resolution of the detector deteriorates. 
For the inverted hierarchy case too, we can see a 
big reduction in the sensitivity due to the systematic uncertainties and the trend with the 
resolution functions is the same. Amongst the 
5 systematic uncertainties that we have included in our analysis, the ones which 
bring about the most impact are the theoretical uncertainties in the atmospheric 
neutrino fluxes coming from the uncertainty in the energy spectrum (called "tilt error" above) 
and from the uncertainty in the zenith angle.  The uncertainties do not cancel between 
different bins and brings about a rather drastic decline in the sensitivity of PINGU to the 
$\oct$.

\section{\label{sec:octcombined}Octant Sensitivity of PINGU Combined with Others Experiments }

Sizable octant sensitivity is expected from the current accelerator and reactor experiments. 
In the light of large $\theta_{13}$, the reach of the combined data from T2K and 
NO$\nu$A towards the discovery of the 
true octant of $\theta_{23}$ has been studied before in \cite{minakataoctant,mahn,tenyrs09} 
and more recently in \cite{t2k2.5,animesh}.
The configuration for NO$\nu$A given in the earlier NO$\nu$A DPR and used in \cite{tenyrs09} 
was revised last year following the discovery of the relatively large $\theta_{13}$. 
The details of the modified configuration for NO$\nu$A can be found in \cite{newnova}
and was used in \cite{skanova} to show the significant improvement expected in the 
mass hierarchy sensitivity of NO$\nu$A. This revised configuration was used 
in the context of $\oct$ sensitivity from current long baseline experiments in 
\cite{t2k2.5}.
In \cite{t2k2.5}, the authors also studied the impact of running T2K for 2.5 years in the 
neutrino and 2.5 years in the antineutrino mode, on the $\oct$ sensitivity. 
The T2K experiment is officially scheduled to run only in the neutrino mode for 5 years. 
However, efforts are on to determine optimized exposure times for T2K in 
neutrino mode vis-o-vis antineutrino mode. On the other hand,
the reactor experiments Daya Bay, RENO and Double Chooz, are expected to 
tighten the constraint on $\theta_{13}$, which as we had seen in the previous 
section, makes a significant difference to the octant sensitivity of PINGU. 
Accurate measurement of $\theta_{13}$ at the reactor experiments is also 
known to be crucial for the octant determination using the T2K and NO$\nu$A experiments. 
In this section we will include the full 3 years run of each of the reactor experiments, 
Daya Bay, RENO and Double Chooz, and combine this with long baseline data 
from T2K and NO$\nu$A. For NO$\nu$A we consider 3 years of neutrino and 3 years of 
antineutrino run and compare results between the old and 
the revised configurations of this experiment, while for T2K we will show octant 
sensitivity plots for both 5 years running in only the neutrino mode and 2.5 years in 
neutrino and 2.5 years in the antineutrino mode. 
Simulations of the event rates and $\Delta \chi^2$ for the long baseline and reactor 
experiments have been performed using the GLoBES software \cite{globes}. 
Finally, we will combine the 
reactor and long baseline data with 3 years of PINGU data and present the 
$\oct$ sensitivity expected from the global data set. 
\\

\begin{figure}
\centering
\includegraphics[width=0.495\textwidth]{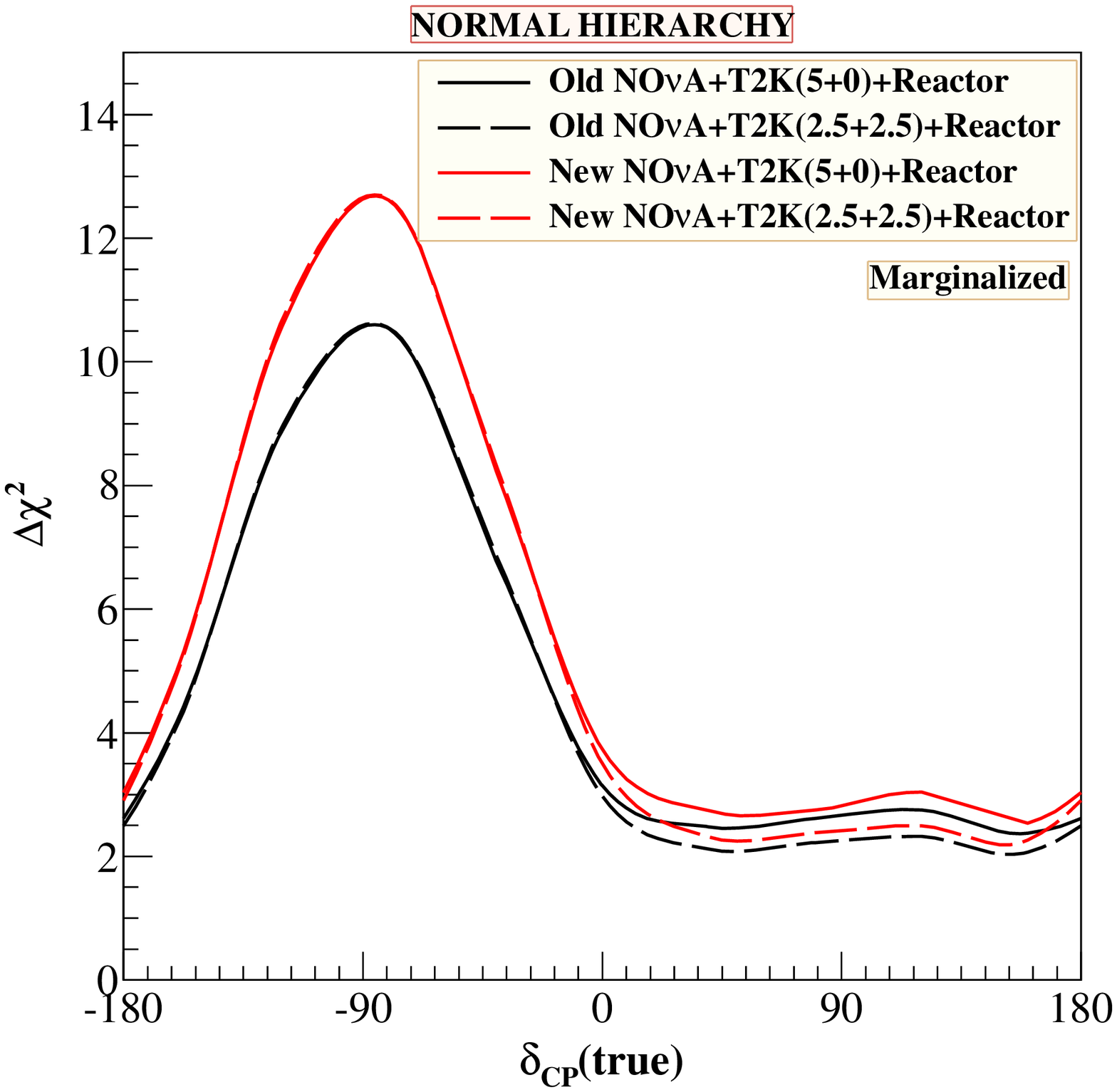}
\includegraphics[width=0.495\textwidth]{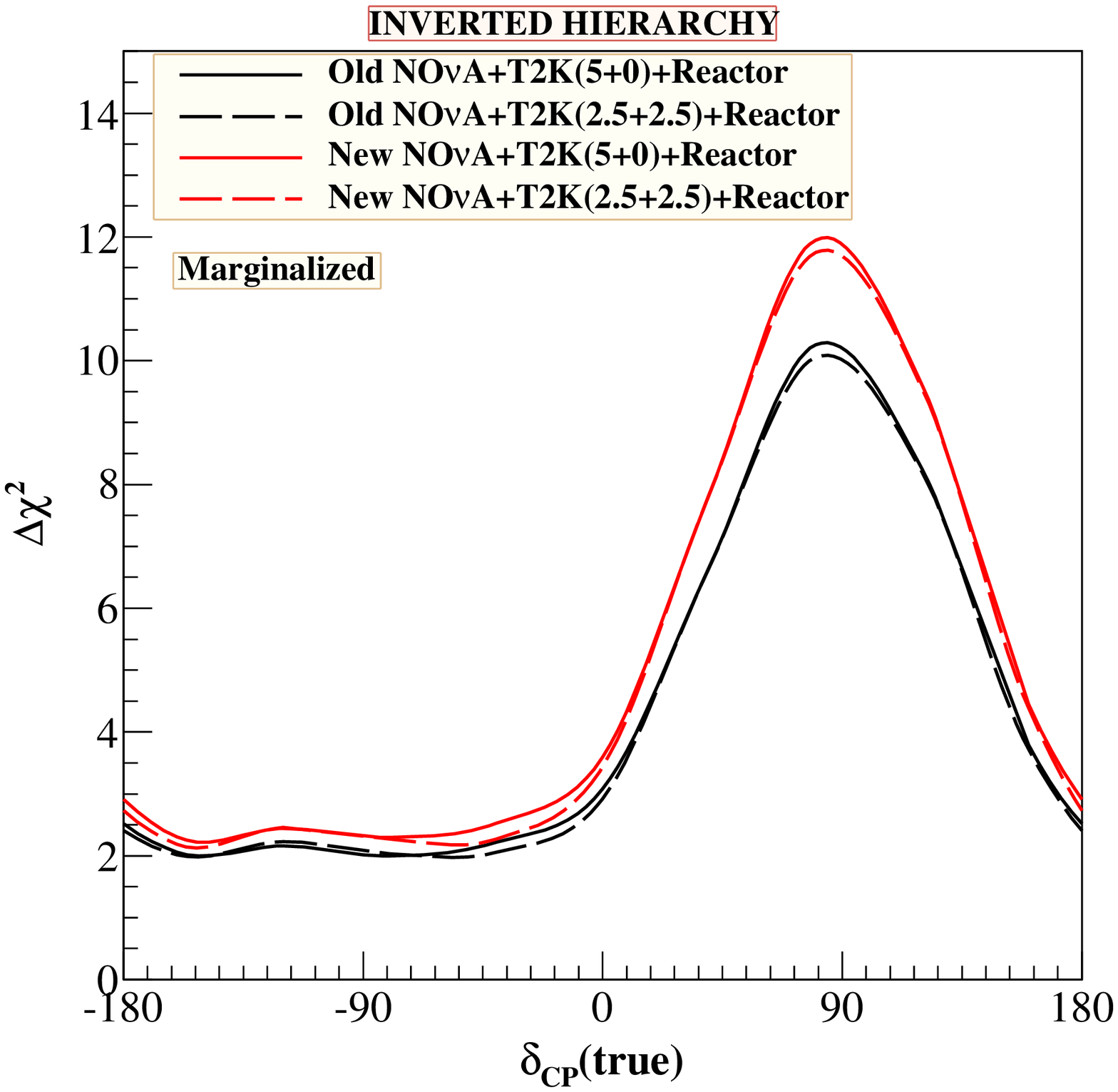}
\caption{\label{fig:lblmh}
The $\Delta \chi^2$ for 
mass hierarchy sensitivity as a function of $\dcpt$ for different 
configurations of NO$\nu$A and T2K. For all cases we have the full projected 
data set from Daya Bay, RENO and Double Chooz reactor experiments. 
The $\Delta \chi^2$ is fully marginalized over the oscillation parameter region
as described in the Table \ref{tab:param}. 
}
\end{figure}

In Fig. \ref{fig:lblmh}
we show the impact of using the revised NO$\nu$A configuration 
on the mass hierarchy sensitivity from the combined analysis of the projected 
data from T2K, NO$\nu$A, Daya Bay, RENO and Double Chooz. 
The $\Delta \chi^2$ for the mass hierarchy sensitivity with the new NO$\nu$A configuration 
is seen to agree well with a similar plot shown in \cite{skanova}, apart from the 
fact that we have used $\stcht=0.1$ for this plot while the $\stcht$ used in 
\cite{skanova} was slightly lower. The other difference is that we have explicitly 
added the reactor data to constrain $\theta_{13}$ while the authors of 
\cite{skanova} impose a prior on $\stch$. The mass hierarchy sensitivity 
is seen to increase with the new NO$\nu$A configuration. 
We also show in this plot the impact of running T2K for 2.5 years in the  
neutrino and 2.5 years in the antineutrino mode. The solid lines are for 5 years of 
T2K in only neutrino mode, while the dashed lines are with the 2.5+2.5 years configuration. 
The mass hierarchy sensitivity is known to be almost independent of the addition of  
T2K data in the $\dcpt=[-\pi,0]$ range for true normal hierarchy
and $\dcpt=[0,\pi]$ range for true inverted hierarchy, where NO$\nu$A gives very high sensitivity to the 
mass hierarchy. However, in the region $\dcpt=[0,\pi]$ for true normal hierarchy and the region 
$\dcpt=[-\pi,0]$ for the true inverted hierarchy, the T2K data does play a major role in 
increasing the mass hierarchy sensitivity of the combined analysis synergetically. 
In this relevant region, it is seen from the figure that running T2K for 2.5 years in neutrino 
and 2.5 years in antineutrino mode will result in a small loss in the net mass hierarchy 
sensitivity. 
\\

\begin{figure}
\centering
\includegraphics[width=0.495\textwidth]{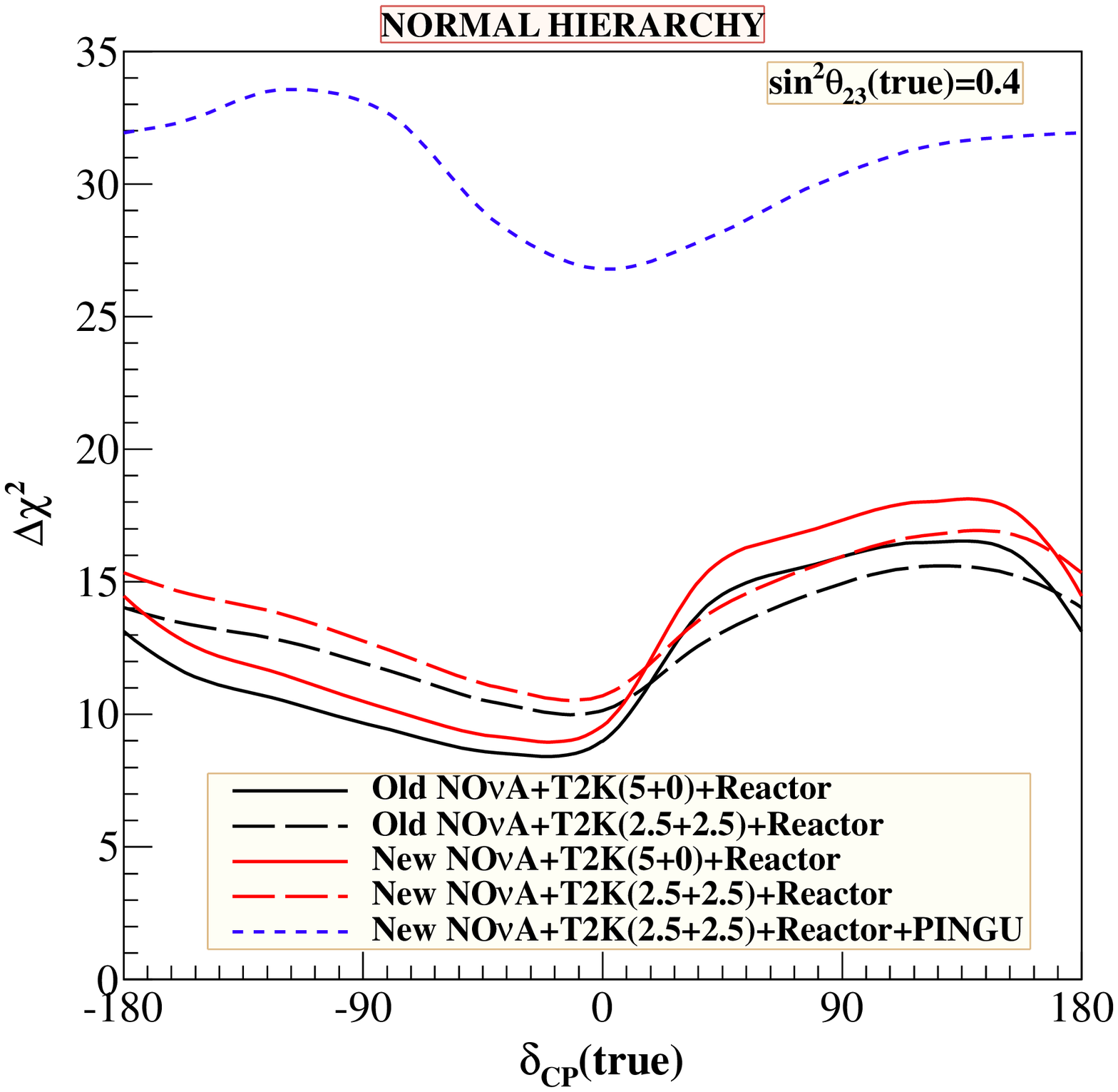}
\includegraphics[width=0.495\textwidth]{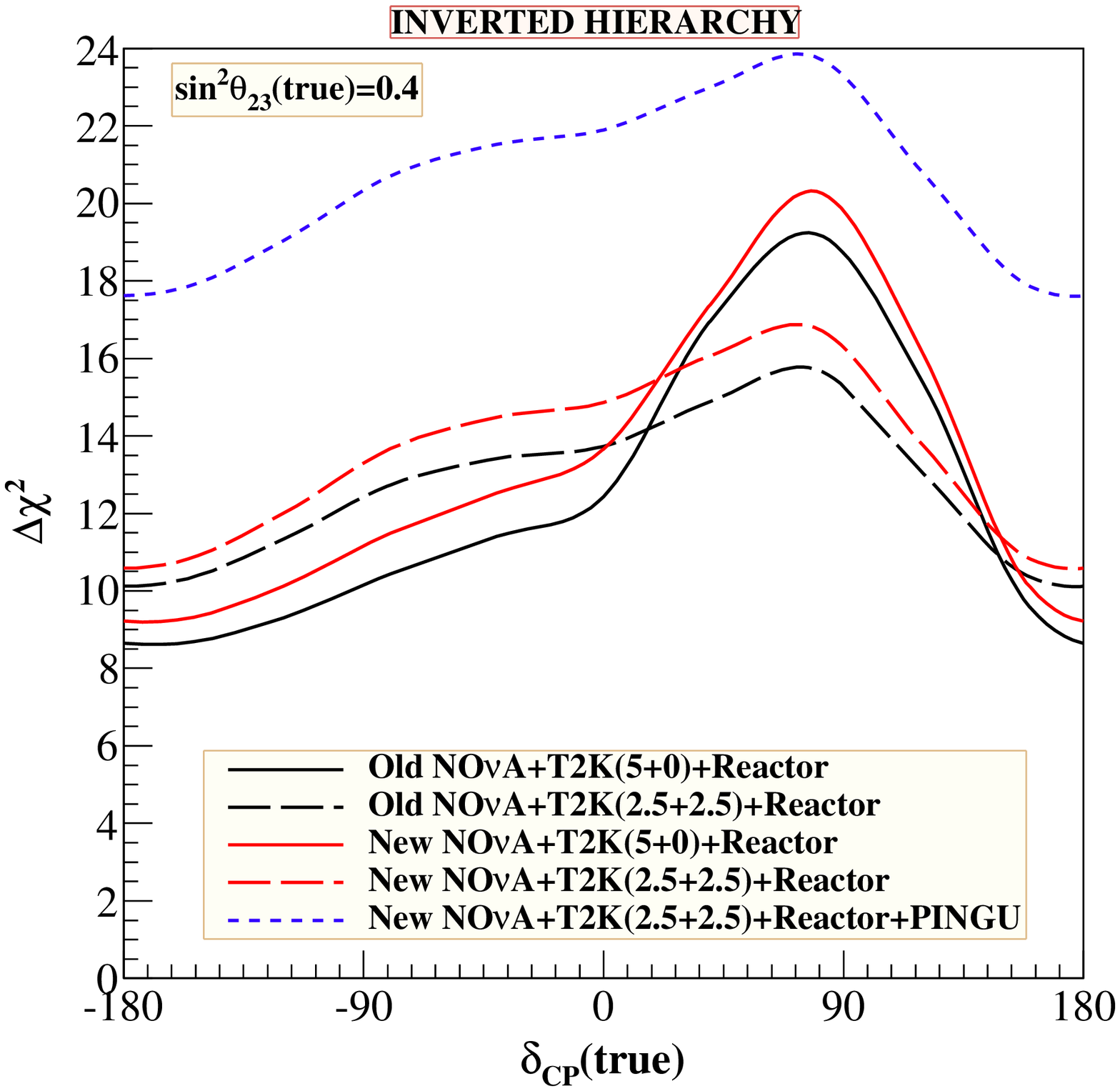}
\includegraphics[width=0.495\textwidth]{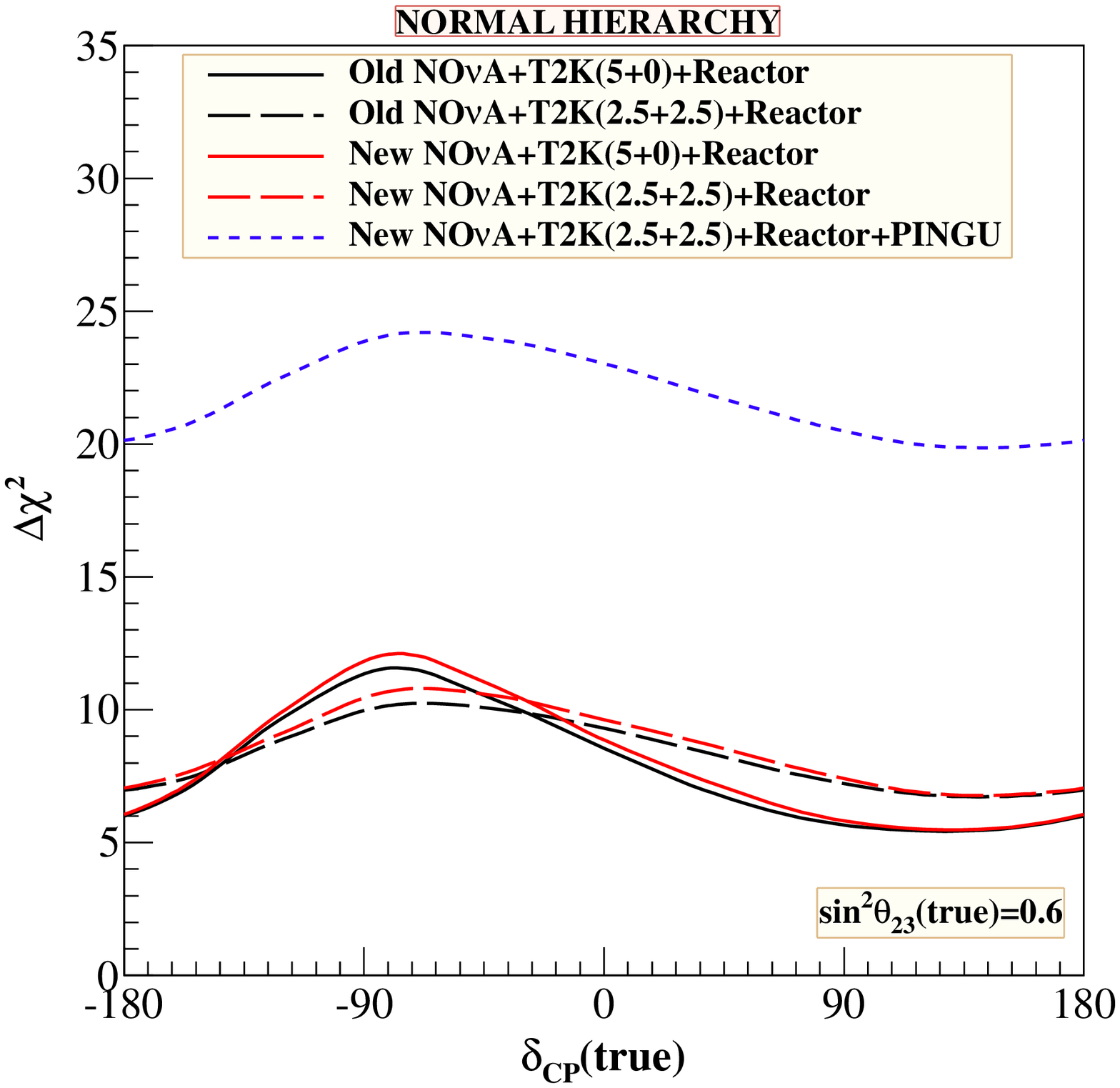}
\includegraphics[width=0.495\textwidth]{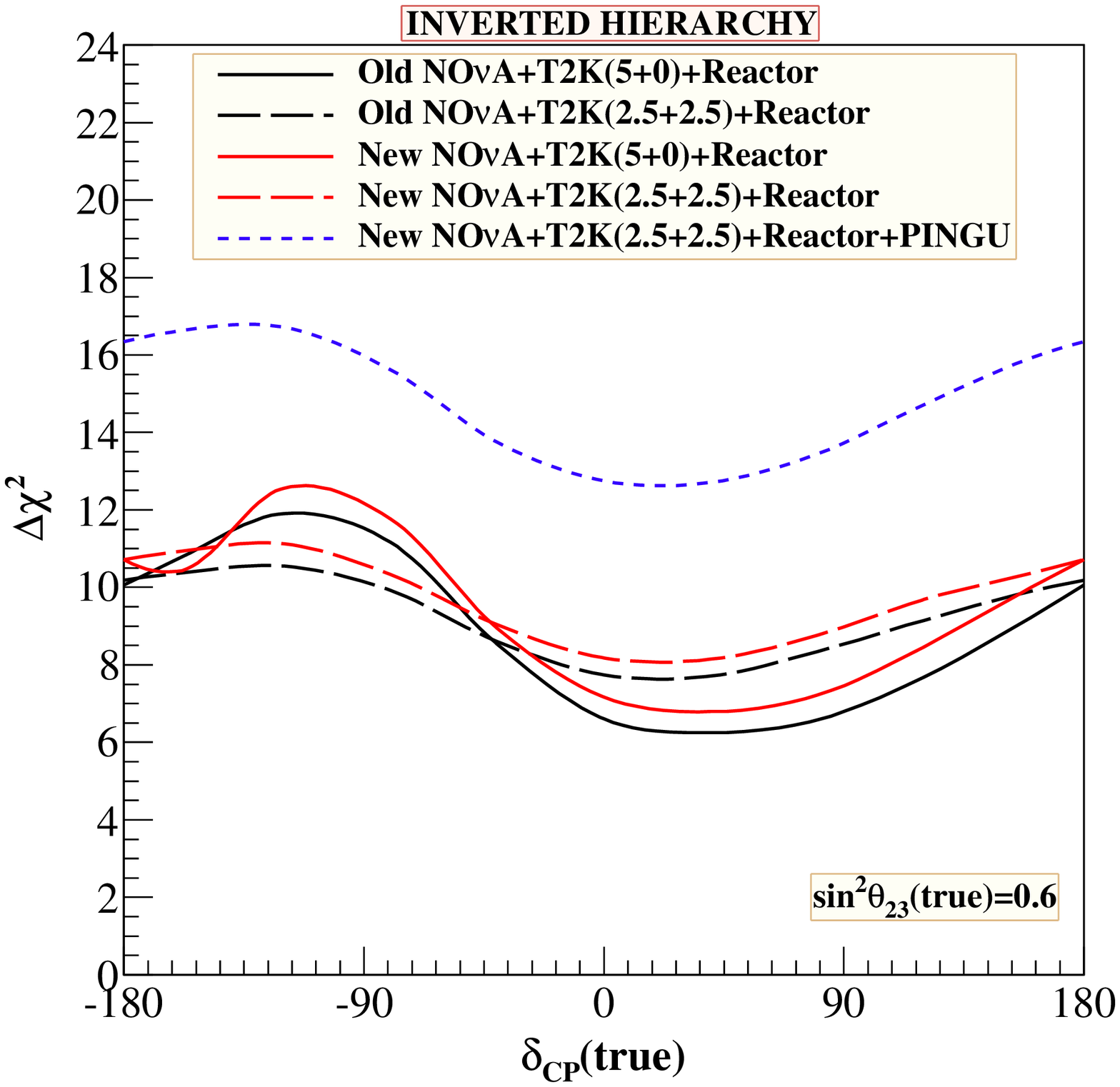}
\caption{\label{fig:lbloctant}
The $\Delta \chi^2$ for the octant sensitivity as a function of $\dcpt$ for different 
configurations of NO$\nu$A and T2K. For all cases we have the full projected 
data set from Daya Bay, RENO and Double Chooz reactor experiments. 
The $\Delta \chi^2$ is fully marginalized over the oscillation parameter region
as described in the Table \ref{tab:param}, as well as the mass hierarchy. The upper panels are 
for $\sat=0.4$ while the lower panels are for $\sat=0.6$. The left-hand panels are for true 
normal hierarchy while the right-hand panels are for true inverted hierarchy. 
}
\end{figure}

The $\dcp$-$\theta_{23}$ degeneracy has been discussed recently in 
\cite{t2k2.5,machado}. It has been shown in these works that this degeneracy can be broken 
if one allows T2K to run partially in the antineutrino mode as well. 
We show the octant sensitivity from the combined analysis of 
NO$\nu$A, T2K and the reactor experiments Daya Bay, RENO and Double Chooz, 
in Fig. \ref{fig:lbloctant}. The figure shows the $\Delta \chi^2$ for the wrong octant 
solution as a function of $\dcpt$. The left-hand panels are for true normal hierarchy 
while the right-hand panels are for true inverted hierarchy. We show the octant 
sensitivity for two benchmark values of $\sat=0.4$ (upper panels) and 
$\sat=0.6$ (lower panels). Each of the $\Delta \chi^2$ shown are fully marginalized over 
$|\meff|$, $\sch$, $\sa$, as well as the mass hierarchy. 
The black lines are for the combined analysis of 
the long baseline and reactor data with the old NO$\nu$A configuration, while the 
red lines are with the new NO$\nu$A configuration. The red lines in the Fig. \ref{fig:lbloctant}
are always above the black lines for both hierarchies and for both the benchmark 
$\sat$ chosen. Therefore, the new NO$\nu$A configuration is seen to perform better 
for the octant sensitivity as well. The comparison between the T2K set-up running 
for full 5 years in the neutrino mode alone (solid lines) 
with that where T2K is run for 
2.5 years in neutrino and 2.5 years in the antineutrino mode (dashed lines)
is also shown in this figure. While the T2K 2.5+2.5 years set-up is seen to reduce the 
$\Delta \chi^2$ for 
$\dcpt$ values where the octant sensitivity of the combined long baseline and 
reactor data was higher, it increases the $\Delta \chi^2$ in the 
$\dcpt$ range where the sensitivity was lower. In particular, breaking up of the 
T2K running time from 5 years in neutrino to 2.5+2.5 years in neutrino+antineutrino mode 
reduces the fluctuation of $\Delta\chi^2$ as a function of $\dcpt$. 
\\

The blue short-dashed lines in Fig. \ref{fig:lbloctant} are obtained when the 
T2K, NO$\nu$A and the reactor data are combined with 3 years of data from the 
PINGU atmospheric neutrino experiment. We have used the PINGU configuration 
where $\sigma_E = 0.2E'$ and $\sigma_\Theta = 0.5\,(1~{\rm GeV}/\sqrt{E'})$ and have 
included the 5 systematic uncertainties as discussed before. 
The PINGU data has almost no dependence 
on $\dcpt$ and adds a constant $\Delta \chi^2$ constant contribution to the octant 
sensitivity of the global data. However, note that this additive contribution is more than 
what we were getting from PINGU alone. The increase in the $\Delta \chi^2$ from 
PINGU comes from the constraint on $\theta_{13}$ imposed by the reactor data and 
from the constraint on $\meff$ and $\sta$ from the long baseline data. This results in 
reducing the effect of marginalization over these parameters. This shows the 
synergy between PINGU and the long baseline and reactor data. 
If we were to add any further data set 
which could help in resolving the neutrino mass hierarchy, like the data from the 
INO atmospheric neutrino experiment \cite{mh}, then this could further lift the 
$\Delta \chi^2$ for octant determination, for the low octant case ($\sat=0.4$) and for 
true normal hierarchy. For the high octant case for true normal hierarchy and for all cases for 
true inverted hierarchy, we had seen in the previous section that marginalization over the 
hierarchy made no difference and hence for these cases information on the neutrino mass 
hierarchy is not expected to bring any further improvement. We must stress here that 
addition of the INO data in the combined analysis will also improve the octant 
determination reach as INO itself has some octant sensitivity. However, we are 
not considering that possibility in this paper and this work will appear elsewhere \cite{oct}. 
For the true normal hierarchy, 
from the combined data from T2K, NO$\nu$A, reactors and 3 years of PINGU, 
we expect a $\Delta \chi^2$ for the wrong octant solution between 
$\Delta \chi^2=26.8-31.9$  for $\sat=0.4$, depending on the value of $\dcpt$ and between 
$\Delta \chi^2=19.9-24.1$  for $\sat=0.6$, depending on the value of $\dcpt$. 
On the other hand, for the true inverted hierarchy, 
from the combined data from T2K, NO$\nu$A, reactors and 3 years of PINGU, 
we expect a $\Delta \chi^2$ for the wrong octant solution between 
$\Delta \chi^2=17.6-23.8$  for $\sat=0.4$, depending on the $\dcpt$ and between 
$\Delta \chi^2=12.8-16.7$  for $\sat=0.6$ depending on the $\dcpt$. 
\\

\begin{figure}
\centering
\includegraphics[width=0.495\textwidth]{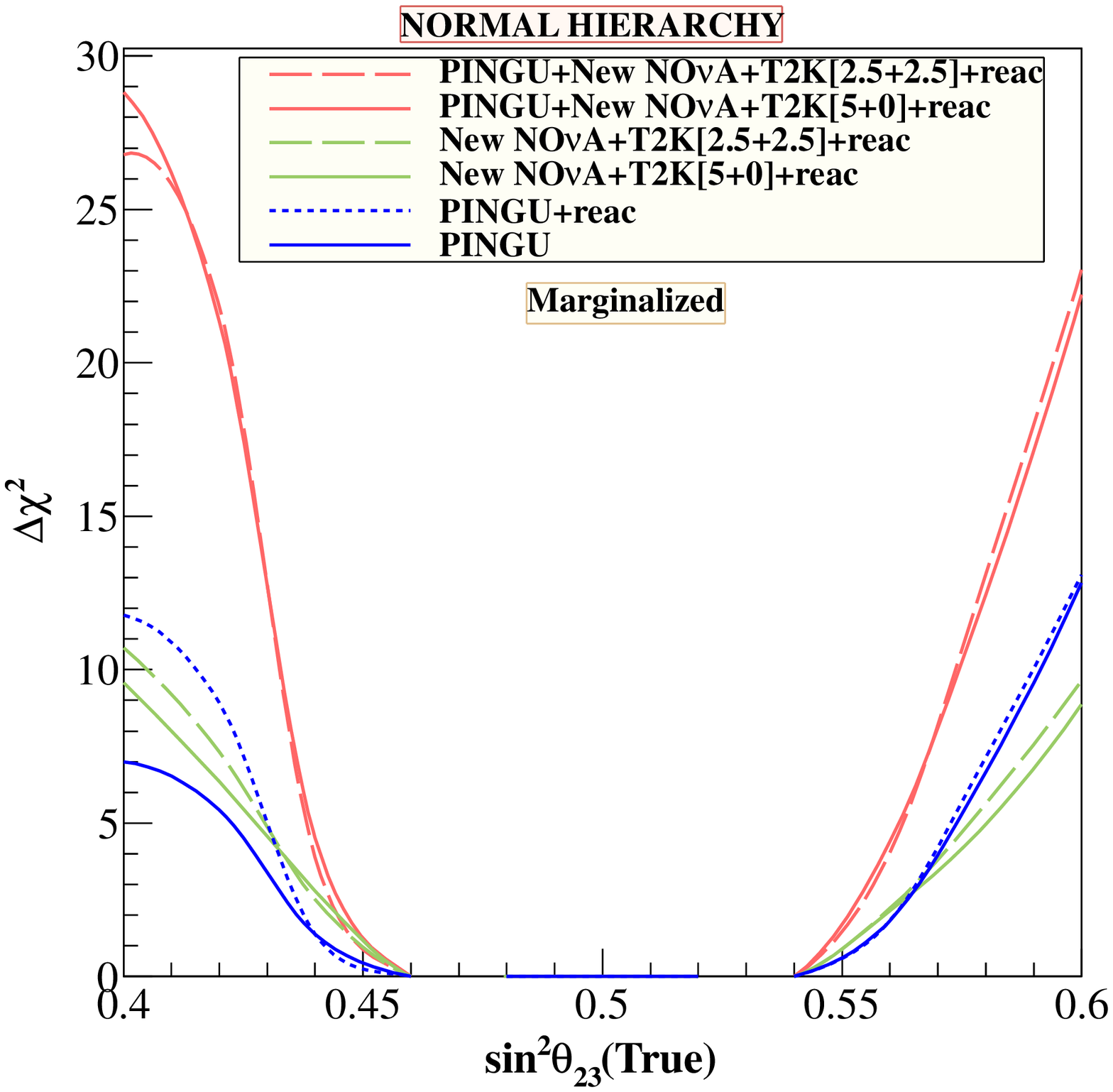}
\includegraphics[width=0.495\textwidth]{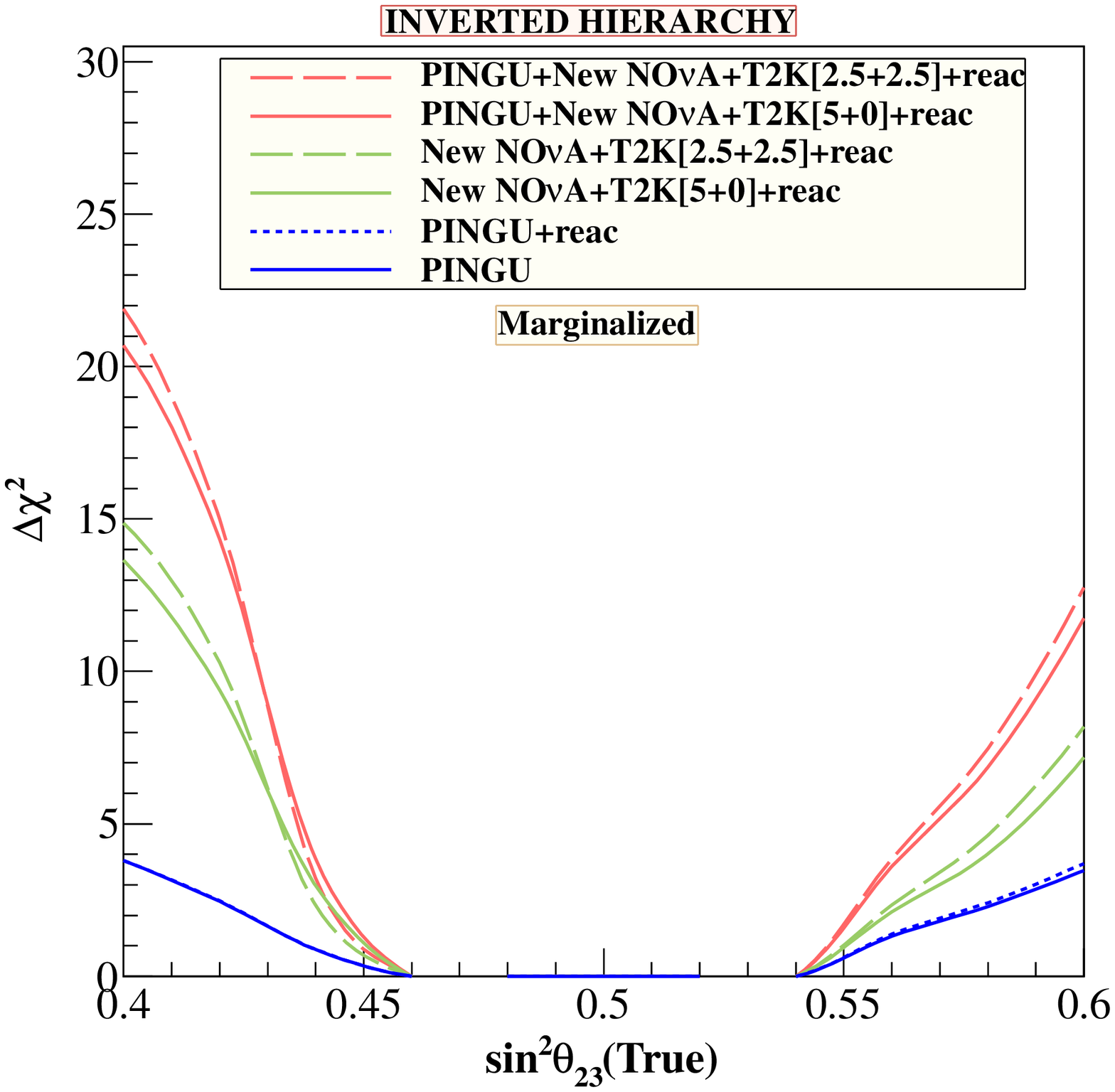}
\caption{\label{fig:comb}
$\Delta \chi^2$ for the wrong octant obtained from the combined analysis of 
3 years of PINGU data with 3+3 years of running of (new) NO$\nu$A, 5+0 years OR 2.5+2.5 years 
of running of T2K and 3 years of running of Daya Bay, RENO and Double Chooz. We show the 
combined sensitivity as a function of $\sat$ for $\dcpt=0$. }
\end{figure}

The effect of combining the long baseline and reactor data with 3 years of PINGU data and 
the final combined sensitivity of the whole set-up together is shown in Fig. \ref{fig:comb}, 
as a function of $\sat$. We choose $\dcpt=0$ to simulate the data for this figure. 
The left-panel of the figure is for true normal hierarchy while the right-panel is for true inverted 
hierarchy. Each of the curves shown in this figure have been fully marginalized in the 
way discussed before. The blue solid line shows the octant sensitivity of PINGU alone 
after marginalizing over oscillation parameters, including the neutrino mass hierarchy. 
Here we take $\sigma_E=0.2E'$ and 
$\sigma_\Theta=0.5\sqrt{E'}$ for PINGU, and full systematic uncertainties as discussed 
before. The blue dotted lines show the $\oct$ sensitivity from 3 years of PINGU when the 
reactor data from 3 years of running of Daya Bay, RENO and Double Chooz are included 
in the analysis. The addition of the reactor data constrains the $\theta_{13}$ range, killing the 
effect of marginalization over this parameter and hence the 
$\Delta \chi^2$ increases for the low values of $\sat$. For the high 
octant side for true normal hierarchy and for all values of $\sat$ for the true 
inverted hierarchy, we do not see any improvement from the restriction on $\theta_{13}$ 
due the reactor data, as was discussed in the 
previous section (cf. Fig. \ref{fig:senspingu_marg}).
\\

The solid green lines are what we expect from T2K+NO$\nu$A+Reactor experiments, where 
we consider 5 years of running of T2K in the neutrino mode alone, while the dashed green 
lines are for the case when T2K is run for 2.5 years in the neutrino mode and 2.5 years in the 
antineutrino mode.
A comparison of the potential of PINGU versus the LBL+Reactor data shows that 
for the true normal hierarchy case, just 3 years of data from PINGU alone 
has a sensitivity to the octant of $\theta_{23}$ comparable to that of the 
combined long baseline and 
reactor data. For values of $\sat$ 
closer to maximal mixing, the PINGU alone analysis suffers mainly from marginalization over 
$\theta_{23}$ in the fit, while for values far away from maximal mixing, especially on the 
low octant side, the PINGU sensitivity 
is constrained mainly from the uncertainty in $\theta_{13}$ and the neutrino mass hierarchy. 
This was demonstrated earlier 
in Fig. \ref{fig:senspingu_marg}. As a result, with the reactor data added the sensitivity 
from PINGU improves further for these values of $\sat$ and for true normal hierarchy. 
Adding the long baseline data helps constrain $\sta$ and $\meff$ better. 
This synergy results in better sensitivity from the combined analysis of 
PINGU with T2K, NO$\nu$A and reactor experiments, and the combined $\Delta \chi^2$ is 
higher than the simple sum of the $\Delta \chi^2$ from PINGU and the long baseline and reactor 
experiments. The combined $\Delta \chi^2$ is shown by the red lines in the figure. The solid 
red line is for T2K running for 5 full years in the neutrino mode while the dashed red lines are for 
T2K running for 2.5 years in the neutrino mode and 2.5 years in the antineutrino mode. 
For normal hierarchy, the combined data set including 3 years of PINGU will return a 
$\Delta \chi^2 > 16$ if $\sat <0.426$ and $>0.586$, and $\Delta \chi^2 > 25$ if $\sat <0.413$. 
For the inverted hierarchy, as discussed before, the octant sensitivity of PINGU is 
low. On combining with the T2K, NO$\nu$A and reactor data we expect 
$\Delta \chi^2 > 9$ if $\sat <0.43$ and $>0.585$, and $\Delta \chi^2 > 16$ if $\sat <0.417$. 
\\

Finally, we have checked that the $\Delta \chi^2$ for the wrong octant from PINGU grows 
almost linearly with the number of years of running of the experiment. We have presented all results 
in this paper for a very short running time of only 3 years. But as this experiment accumulates 
more data, the statistical significance with which if can diagnose the true octant of $\theta_{23}$ 
will grow.

\section{\label{sec:concl}Conclusions}

In this paper we explored the prospects of determining the octant of $\theta_{23}$ 
with atmospheric neutrinos at PINGU. For illustration, we simulated 3 years of data 
at this experiment with some benchmark detector specifications. We used the
effective volume of the detector read 
from \cite{pingutalk} for the curve labelled ``Triggered Effective Volume, R=100m". For the
energy and angle resolutions of the neutrino, we used some benchmark values  
consistent with that used in the literature. We used Gaussian function for the resolution functions with 
our optimal choice for the width of these Gaussians being 
$\sigma_E = 0.2E'$ and $\sigma_\Theta = 0.5\,(1~{\rm GeV}/\sqrt{E'})$ for the 
energy and angle resolution, respective. We studied the impact of the worsening of the 
resolutions, both in energy and angle. We defined a $\chi^2$ and showed the statistical 
significance with which the octant of $\theta_{23}$ can be determined with just 3 years of 
running of PINGU. In our analysis we included 5 systematic uncertainties, 4 of which 
come from uncertainties in the theoretical fluxes. 
We showed in detail the impact of systematic uncertainties on the final $\chi^2$ from PINGU. 
The systematic uncertainties were seen to make a huge impact on the sensitivity, with the 
$\chi^2$ reducing by as much as a factor of 3.5 to 7.5, depending on the energy and angle 
resolutions chosen.
Therefore, we conclude that the inclusion of systematic uncertainties on the theoretical 
fluxes is a must for any PINGU analysis.
\\

We also looked at the impact of the uncertainties on the neutrino oscillation parameters. 
We presented results where the wrong octant $\chi^2$ from PINGU alone was marginalized 
over all oscillation parameters, including the neutrino mass hierarchy. The effect of marginalizing 
over $\theta_{23}$ was seen to be dominant for both true normal as well as true inverted hierarchy. 
The effect of $\theta_{13}$, $|\meff|$ and the neutrino mass hierarchy was seen to be 
dominant for the true normal hierarchy and lower values of $\sat$. The inclusion of the prospective  
3 years of running of the Daya Bay, RENO and Double Chooz reactor data was shown to 
alleviate the problem from parameter uncertainties in this region by severely constraining the 
allowed range of $\theta_{13}$. 
We found that 
for the true normal hierarchy, with just 3 years of 
PINGU data {\it alone}, one could rule out the wrong hierarchy at the $2\sigma$ C.L. 
for $\sat<0.427$ and $>0.569$. With addition of the data from the reactor experiments, 
we could achieve $3\sigma$ C.L. determination of the octant for 
$\sat<0.419$ and $>0.586$ for the true normal hierarchy. The sensitivity of PINGU to the 
octant of $\theta_{23}$ is shown to be rather low for the true inverted hierarchy. The 
$\Delta \chi^2$ for the wrong hierarchy goes linearly with the statistics in this regime. Hence 
the sensitivity will go up as we increase the number of year of running of PINGU. 
\\

We also looked in detail at the prospects of measuring the octant of $\theta_{23}$ at the 
long baseline experiments T2K and NO$\nu$A in conjunction with the reactor data which 
synergetically determines the value of $\theta_{13}$ in a clean way. 
The NO$\nu$A configuration has been re-optimized by the collaboration following the 
measurement of $\theta_{13}$. Though the T2K experiment is still slated to run only in the 
neutrino mode for the full 5 years, there have been recent studies looking to re-optimize the 
T2K run by introducing a few years of antineutrino running for this experiment.  In this paper, 
we looked at the physics reach of both old as well as new NO$\nu$A, and T2K run in 
both the 5 years in the neutrino mode and 2.5 years of neutrino and 2.5 years in the 
antineutrino mode. 
We started by comparing the impact on 
the mass hierarchy sensitivity due to the change of the NO$\nu$A and T2K configurations. 
We showed that while the new NO$\nu$A configuration improves the mass 
hierarchy sensitivity for the entire range of $\dcpt$, running the T2K in the 2.5+2.5 years 
of neutrino plus antineutrino mode results in a small reduction of the mass hierarchy sensitivity. 
We next looked at the octant sensitivity from these experiments along with the reactor 
experiments, and again we made a comparative study of the new vs old NO$\nu$A and 
T2K. As pointed out in the literature before, we found that the addition of the antineutrino run 
in the T2K data helps in cancellation of the octant-$\dcp$ degeneracy. As a result, in regions of 
$\dcpt$ where the combined sensitivity of T2K, NO$\nu$A and the reactors to the octant is 
low, the addition of the antineutrino run improves the sensitivity to the $\oct$. 
\\

Finally, we added 3 years of projected PINGU data to the projected data from T2K, NO$\nu$A and 
the reactors and showed the results both as a function of $\dcpt$ as well as $\sat$. Three years of 
PINGU data along with data from T2K, NO$\nu$A and the reactors will be able to pin down the 
right octant of $\theta_{23}$ for the true normal hierarchy 
to better than $4\sigma$ C.L. for all values of $\sat<0.426$ and $>0.586$.  
A $5\sigma$ significance for the right octant can be achieved if $\sat<0.413$ 
for the true normal hierarchy. If the inverted hierarchy were true, then we would have a 
$3\sigma$ determination of $\oct$ for $\sat<0.43$ and $>0.585$ from the combined data set 
with of 3 years of PINGU. The sensitivity will scale as we increase the number of years of 
running of the PINGU experiment and the $\Delta \chi^2$ coming from PINGU increases 
almost linearly with the number of years of running of this experiment.

\vskip 1cm
\noindent
{\Large \bf Acknowledgements}
\\

We thank S. Prakash for discussions. 
S.C. acknowledges partial support from the European Union FP7 ITN INVISIBLES
(Marie Curie Actions, PITN-GA-2011-289442).


\end{document}